\documentclass[showpacs,preprintnumbers,amsmath,amssymb,footinbib]{revtex4-1}
\usepackage{graphicx}
\usepackage{epsfig}
\usepackage{color}
\usepackage{subfig}
\usepackage[english]{babel}
\usepackage{amsmath,amssymb,amsfonts}
\usepackage{dcolumn}
\usepackage{bm}
\usepackage{mathrsfs}
\usepackage{multirow}
\usepackage[table]{xcolor}
\begin{document}

\title{ Spin effects on the semiclassical trajectories of Dirac electrons}
\author{R. Guti\'errez-J\'auregui$^1$, R. P\'erez-Pascual$^{2}$, R. J\'auregui$^2$
}
\affiliation{$^1$ The Dodd-Walls Centre for Photonic and Quantum Technologies, Department of Physics, University of Auckland,
Private Bag 92019, Auckland, New Zealand\\ $^2$ Instituto de F\'{\i}sica, Universidad Nacional Aut\'onoma de M\'exico, Apdo. Postal 20-364, 01000 Cd. de M\'exico, M\'exico}

\begin{abstract}The relativistic semiclassical evolution of the position of an electron in the presence of an external electromagnetic field is studied in terms of a Newton equation that incorporates spin effects directly.
This equation emerges from the Dirac equation and allows the identification of scenarios where spin effects are necessary to understand the main characteristics of the electron  trajectories.
It involves the eigenvalues of the non-Hermitian operator $\Sigma_{\mu\nu}F^{\mu\nu}$ with $\Sigma_{\mu\nu}$ and $F^{\mu\nu}$ as the spin and electromagnetic tensors. The formalism allows a deeper understanding on the physics behind known analytical solutions of the Dirac equation when translational dynamics decouples from spin evolution. As an illustrative example, it is applied to an electron immersed in an electromagnetic field which exhibits chiral symmetry and optical vortices. It is shown that the polarization of intense structured light beams can be used to suppress or enhance spin effects on the electron semiclassical trajectory;  the latter case yields a realization of a Stern-Gerlach apparatus for an electron.
\end{abstract}
\maketitle

\section{Introduction}
Up to radiative corrections, an electron in an external electromagnetic (EM) field is described with high accuracy by the Dirac equation. A few years after Dirac's proposal, Pauli pioneered the analysis of semiclassical limits of this equation \cite{pauli}.
Pauli noticed that an approach based on taking Planck constant to a zero value directly eliminates spin effects.
Along this line of thought, Rubinow and Keller \cite{RBK} worked out a formalism  where  the electron
semiclassical trajectories are determined by the Lorentz force, while the electron spin precesses
as described by the Bargmann-Michel-Telegdi (BMT) equation \cite{BMT}.
In several physical configurations, such a description is more reliable than one could expect. For instance,
the motion of an electron located in  a close realization of an EM plane wave is not influenced by its
spin state: the known exact solution of the Dirac equation predicts the uncoupling
between spin and translational states  \cite{Volkov1935,bagrov}. Another  example is provided by the experimental arrangement for the measurement of the gyromagnetic ratio  of the electron.
It is based on the  comparison of  the frequency of the electron spin precession with the cyclotron frequency of its orbit,  both due to the presence of a static magnetic field  \cite{dehlmelt,gabrielse}.
The results are roughly described taking both dynamical variables as uncoupled; that is due to the high strength of the Lorentz interaction with respect to the interaction of a particle with
magnetic dipole moment given by the Bohr magneton $\mu_B$.

Forthcoming experimental efforts involving electrons in the presence of spatially focused intense laser beams \cite{focused} could give rise to physical arrangements
that break this approach.
In these kinds of systems, the large energy of the electron allows a description that, to a first approximation, can be based on semiclassical worldlines
from which relevant quantum states can be identified via a WKB formalism \cite{piazza}.  However, in such analyzes
spin effects on the semiclassical trajectories have not yet been considered and, as we shall show, could be important.

In this article, we  obtain an equation for the trajectory of  electrons that incorporates  spin effects directly.
It emerges from the Dirac equation and clearly shows the dominant role of the Lorentz force
in situations as those described above; it also allows the identification of scenarios where spin effects
are necessary to understand the main characteristics of the electron   trajectories.

In general, successful approximate approaches to the Dirac equation
require the usage of a basis set of spinors that incorporate the essential features of the physical configuration under study \cite{Drake,Jauregui}.
Here, we shall take a basis that is derived from eigenspinors of the non-Hermitian operator $\Sigma_{\mu\nu}F^{\mu\nu}$, with $\Sigma_{\mu\nu}$ and $F^{\mu\nu}$ as the spin and EM tensors respectively.
This operator appears in the Dirac-Pauli equation; solutions of the Dirac equation are obtained from the solutions of Dirac-Pauli equation via a differential operator $\hat{\mathcal{O}}_D$.
The complex eigenvalues   of $\Sigma_{\mu\nu}F^{\mu\nu}$ have  direct physical interpretations. In fact, these eigenvalues allow a deeper understanding on the physics behind known analytical solutions of the Dirac equation in the presence of EM waves where translational dynamics seem to decouple from spin evolution \cite{Volkov1935,birula}.
The Dirac spinors that result from applying the operator $\hat{\mathcal{O}}_D$ to the eigenspinors of $\Sigma_{\mu\nu}F^{\mu\nu}$ are shown to have
a simple structure which involves essential features of the EM field, $e.$ $g.$, the current of its source and the EM energy and momentum densities.
The Dirac-Pauli equation is written in terms of the eigenspinors and eigenvalues of $\Sigma_{\mu\nu}F^{\mu\nu}$.
We then proceed to incorporate such eigenvalues in the Newton equations that should be satisfied by the semiclassical trajectories of Dirac electrons.
A  self consistent procedure to obtain approximate solutions of the Dirac-Pauli equation (which yield approximate solutions of the Dirac equation) is described.
Finally, a particular example is worked out to illustrate  the formalism, and to show a feasible scenario where spin effects on the semiclassical trajectories are not masked out by the Lorentz force.

\section{Dirac-Pauli equation: a suitable basis set}
The Dirac equation describes an electron as a point particle of mass $m$,  electric charge $q$ and  dipole magnetic moment  $\mu_B = q\hbar/2mc$,
that depends on Planck constant $\hbar$ and the velocity of light $c$.
In the presence of an external EM field with potential $A_\mu$,
any solution $\Psi$ of the Dirac equation also satisfies the Dirac-Pauli relativistic second order equation
\begin{equation}
\Big[(\hat p_\mu-\frac{q}{c} A_\mu)(\hat p^\mu -\frac{q}{c}A^\mu)- \frac{q\hbar}{2c}\Sigma^{\mu\nu}F_{\mu\nu} - m^2c^2\Big]\Psi = 0,
\end{equation}
\begin{equation}
\Sigma^{\mu\nu} =\frac{i}{2}[\gamma^\mu,\gamma^\nu]_-, \quad F_{\mu\nu} = \partial_\mu Ạ_\nu -\partial_\nu A_\mu.
\end{equation}
We work with the metric tensor with signature (1,-1,-1,-1),
and use the representation of the $\gamma^\mu$ matrices
\begin{equation}
\gamma^0 =\begin{pmatrix}  1&0 \\ 0&-1  \end{pmatrix},\quad\quad
\vec \gamma =\begin{pmatrix}  0&-\vec{\mathbb{\sigma}}\\ \vec{\mathbb{\sigma}} &0 \end{pmatrix},
\end{equation}
in terms of the Pauli matrices $\vec{\mathbb{\sigma}}.$
We study solutions $\Psi_P$ of the Dirac-Pauli equation with the structure
\begin{equation}
\Psi_P = \sum_{j} e^{-i\mathcal{S}^{(j)}/\hbar}d_j\psi_j, \label{eq:psiP}
\end{equation}
each $\mathcal{S}^{(j)}$ and $d_j$ are taken as complex scalar functions of the coordinates $x_\mu$ that are assumed to be analytical on $\hbar$.
The  $\psi_j$ bispinors satisfy the eigenvalue equation
\begin{equation}
\Sigma^ {\mu\nu}F_{\mu\nu}\psi_j = \lambda_j(x)\psi_j.  \label{eq:psiq}
\end{equation}
Two eigenfunctions $\psi_i$ and $\psi_j$ with different eigenvalues of the non Hermitean operator $\Sigma^{\mu\nu} F_{\mu\nu}$ may not be orthogonal
under the standard algebraic  product $\psi_j^\dagger\psi_i$. Nevertheless, for $F_{\mu \nu}$  a real tensor,
\begin{equation}
\Sigma^{\mu\nu} F_{\mu\nu}\psi_j = \lambda_j \psi_j \Rightarrow (\lambda_j^* - \lambda_i)\psi_j^\dagger\gamma^0\psi_i = 0.
\end{equation}
In terms of spinors $\phi^{\text{a,b}}_j$, Eq.~(\ref{eq:psiq})  has the structure
\begin{equation}
[\Sigma^ {\mu\nu}F_{\mu\nu}-\lambda_j\mathbf{1}_{4\times 4}]\psi_j = \begin{pmatrix} \mathfrak{A} &i\mathfrak{D}\\ i\mathfrak{D}&\mathfrak{A}  \end{pmatrix}\begin{pmatrix} \phi^\text{a}_j \\ \phi^\text{b}_j \end{pmatrix}= 0,
\end{equation}
\begin{equation}\mathfrak{A} =
\vec\sigma\cdot\vec B -\lambda_j\mathbf{1}_{2\times 2},\quad\quad
\mathfrak{D} =-\vec \sigma\cdot\vec E.
\end{equation}
 Demanding
\begin{equation}
\mathrm{Det}\Big[\Sigma^ {\mu\nu}F_{\mu\nu} - \lambda_j\mathbf{1}_{4\times 4}] =
\mathrm{Det}\Big[\mathfrak{A} +i\mathfrak{D}\Big]\mathrm{Det}\Big[\mathfrak{A} -i\mathfrak{D}\Big]= 0,\end{equation}
implies
\begin{eqnarray}
\lambda_j^2 &=& [\vec B\cdot\vec B - \vec E\cdot\vec E\pm 2i \vec E \cdot \vec B]\nonumber\\
            &=& [\vec G^{\pm}\cdot \vec G^{\pm}],\quad \quad \vec G^{\pm} =: \vec B \pm i\vec E ,\label{eq:ev}
\end{eqnarray}
$\lambda_j$ depends only  on the relativistic invariants
$F_{\mu \nu} F^{\mu \nu} =\vec B\cdot\vec B - \vec E\cdot\vec E =:\Delta^2_{BE}$ and $\epsilon_{\mu \nu \rho \eta} F^{\mu \nu}F^{\rho \eta} =
2\vec E\cdot\vec B$.

 If $ \vec E \cdot \vec B\ne 0$, four different complex eigenvalues are expected, while for
 $ \vec E \cdot \vec B= 0$ a two fold degeneracy arises. The latter condition is fulfilled at least for two very
relevant configurations: (i) a classical Stern-Gerlach experiment with  $\vec E =\vec 0$
in the reference frame where the magnets that generate the magnetic field are at rest, and (ii) an electron moving
in the presence of an EM plane wave in otherwise vacuum. Plane waves are a particular example of the so called crossed fields \cite{bagrov} for
which both invariants are zero for any coordinate $x_\mu$, $\lambda_j \equiv 0$.
Notice, however, that the EM field of structured light beams may satisfy $\vec E\cdot\vec B=0$, but, in general, $\vec B\cdot\vec B \ne \vec E\cdot\vec E$
yielding  eigenvalues  $\lambda_{j}=(-1)^j\sqrt{\Delta^2_{BE}}$ which are either completely real or imaginary in different space regions.

For $ \lambda_j\ne 0$, the eigenvectors $\psi_j$ have the structure
\begin{equation}
\psi_\pm = \begin{pmatrix}\phi^\text{a}_\pm\\\mp\phi^\text{a}_\pm\end{pmatrix}\label{eq:strcture}
\end{equation}
with
\begin{equation}
\phi^\text{a}_\pm = (\vec \sigma\cdot \vec G^{\pm} + \lambda_\pm)\alpha_\pm,\quad  \lambda_\pm^2 =\vec G^{\pm}\cdot\vec G^{\pm};\label{eq:strcture2}
\end{equation}
while for $\vec E \cdot \vec B= 0$ and $\lambda_j\ne 0$ one could alternatively consider,
\begin{eqnarray}
\psi_j &=&\begin{pmatrix} \vec\sigma\cdot\vec B + \lambda_j & 0 \\ 0  &-i\vec\sigma\cdot \vec E \end{pmatrix}\begin{pmatrix} \alpha_1 \\ \alpha_1 \end{pmatrix}
+  \begin{pmatrix} -i\vec\sigma\cdot \vec E  & 0 \\ 0 &\vec\sigma\cdot\vec B + \lambda_j\end{pmatrix}\begin{pmatrix} \alpha_2 \\ \alpha_2\end{pmatrix}.
\end{eqnarray}
In these equations
$\alpha_\pm$, $\alpha_1$ and $\alpha_2$ are arbitrary two component spinors which may depend on spacetime variables.

A procedure to generate sets $\{\psi_j\}$ of four linearly independent bispinors that preserve  the structure described by Eq.~(\ref{eq:strcture}) is presented in Appendices A and B.  The algorithms that generate these basis sets rely on particular selections of the spinors $\alpha_\pm$. Note that Eq.~(\ref{eq:strcture}) automatically guarantees the orthogonality between the bispinors that depend on $\vec G^+$ and those
that depend on $\vec G^-$. So that, for a given index $j$, there is a unique index $i_j$ for which $\psi_j^\dagger\psi_{i_j} $ is non identically zero. For algebraically normalized bispinors $\psi_j$
and $\psi_{i_j}$,
 the bispinors $\Xi_j$ defined by
\begin{equation}
\Xi_j = \frac{\psi_j - (\psi_{i_j}^\dagger\psi_j)\psi_{i_j}}{ 1 -\vert \psi_{i_j}^\dagger\psi_j\vert^2}\label{eq:proj1}
\end{equation}
satisfy the relations
\begin{equation}
\Xi_j^\dagger\psi_i = \delta_{ij},\quad \sum_{i=1}^4 \psi_i\Xi_i^\dagger = \mathbf{1}_{4\times 4} .\label{eq:proj2}
\end{equation}
The properties of the bispinors built in Appendix A is greatly simplified by using the representation of
spinors $\alpha_\pm$ in the Bloch sphere. The dynamical variables of the EM field, such as the density of energy and momentum,
appear in the normalization and overlap of the bispinors in the basis set $\{\psi_j\}$.

\section{Dirac bispinors}
Dirac bispinors are the solutions of the Dirac equation. From them all the quantum properties
of the electron must be extracted.

Dirac bispinors can be  built from  Pauli bispinors since solutions $\Psi_D$ of the Dirac equation
\begin{equation}
\Big[\gamma^\mu (\hat p_\mu -\frac{q}{c} A_\mu) - mc\Big]\Psi_D = 0
\end{equation}
can be obtained from solutions $\Psi_P$ of the Pauli  equation:
\begin{eqnarray}
\Psi_D &=& \Big[\gamma^\mu (\hat p_\mu -\frac{q}{c} A_\mu) + mc\Big]\Psi_P \nonumber\\
       &=:& \hat{\mathcal{O}}_D \Psi_P.\label{eq:DP}
\end{eqnarray}
The structure of the bispinors $\Psi_D$ that result from Eq.~(\ref{eq:psiP}) is simplified by applying Maxwell equations, as we now demonstrate.
We define the vector operator
\begin{equation}
\hat \pi_\mu = \hat p_\mu -\frac{q}{c}A_\mu,
\end{equation}
and the  vector functions
\begin{equation}
\pi_{\mu}^{(j)} = \partial_\mu \mathcal{S}^{(j)} -\frac{q}{c}A_\mu.
\end{equation}
Since the complex functions $ \mathcal{S}^{(j)}= \mathtt{S}^{(j)} + i\mathtt{s}^{(j)}$ have units of action, $\pi_{\mu}^{(j)}$ has units of momentum.
The Dirac bispinor depends on the functions $\pi_\mu^{(j)}$ as follows (see Eq.~(\ref{eq:psiP})),
\begin{eqnarray}
\Psi_D &=& \Big[\gamma^\mu \hat \pi_\mu + mc\Big]\Psi_P\nonumber\\
&=& \sum_j e^{-i\mathcal{S}^{(j)}/\hbar}\Big[\gamma^\mu \pi_\mu^{(j)} + mc + i\hbar\gamma^\mu\partial_\mu\Big]d_j\psi_j.\label{eq:diracterms}
\end{eqnarray}
For $\lambda_j\ne 0$, the $\psi_j$ have the structure shown in Eq.~(\ref{eq:strcture}), that can be synthesized as
\begin{equation}
\psi_j =\begin{pmatrix} \phi^{(j)}_\pm\\ \mp \phi^{(j)}_\pm\end{pmatrix},\quad\quad
 \phi^{(j)}_\pm = \frac{1}{N_j}(\lambda_j +\vec\sigma\cdot\vec G^\pm)\alpha^{(j)}, \label{eq:spinors}
\end{equation}
with $\alpha^{(j)}$ a given spinor.
Let us analyze the behavior of the different terms in Eq.~(\ref{eq:diracterms}).
A direct calculation shows that
\begin{equation}
\Big[\gamma^\mu \pi_\mu^{(j)} + mc\Big]\psi_j = \begin{pmatrix} (\pi_0^{(j)} + mc \pm\vec\sigma\cdot\vec\pi^{(j)}) \phi^{(j)}_\pm\\
\pm(\pi_0^{(j)} - mc \pm\vec\sigma\cdot\vec\pi^{(j)})\phi_\pm^{(j)} \end{pmatrix}.
\end{equation}
Note that although the basis spinor $\psi_j$ presents a balance between its upper spinor $\phi^{(j)}_\pm$ and its lower spinor $\mp\phi^{(j)}_\pm$,
the contribution to the Dirac bispinor $\Psi_D$ of this term breaks the proportionality between the upper and lower spinors through the mass $mc$ factor.
Now, let us obtain the explicit result of applying the operator $\gamma^\mu\partial_\mu$ to each $\psi_j$. The term
\begin{equation}
i\hbar\gamma^\mu\partial_\mu\begin{pmatrix}\vec\sigma\cdot\vec G^\pm\\ \mp\vec\sigma\cdot\vec G^\pm\end{pmatrix} =\begin{pmatrix} i\hbar[\partial_0 \mp \vec \sigma\cdot\vec\nabla]\vec\sigma\cdot\vec G^\pm \\
\mp i\hbar [\partial_0 \mp \vec \sigma\cdot\vec\nabla]\vec\sigma\cdot\vec G^\pm \end{pmatrix},
\end{equation}
is directly related to the charge density $\rho$ and current density $\vec J$ that generate the EM field $F_{\mu\nu}$,
\begin{equation}
 [\partial_0 \mp \vec \sigma\cdot\vec\nabla]\vec\sigma\cdot\vec G^\pm = \vec\nabla\cdot\vec G^\pm + i \vec\sigma\cdot(-i\partial_0 \vec G^\pm + \vec \nabla\times\vec  G^\pm)=
 \pm 4\pi i( \rho \mp \frac{1}{c}\vec\sigma\cdot\vec J)
\end{equation}
as a result of the Maxwell equations. In many of the physical configurations of interest, the
EM field $F_{\mu\nu}$ is external and the electron wave function is negligible in the region where the sources of $F_{\mu\nu}$ are located, then one can take $\rho=0$ and $\vec J = \vec 0$, considerably simplifying the expression of Dirac bispinors $\Psi_D$.
Notice that the factors
\begin{equation}
\Big[\partial_0 \frac{\lambda_j}{N_j}\mp \vec \sigma\cdot\vec\nabla\frac{\lambda_j}{N_j}\Big]\alpha_j +
\Big[\partial_0 \frac{1}{N_j}\mp \vec \sigma\cdot\vec\nabla\frac{1}{N_j}\Big]\vec\sigma\cdot\vec G^\pm\alpha_j
\end{equation}
contained in Eq.~(\ref{eq:diracterms}), involve, by construction, derivatives of both the relativistic invariants $\vert \vec B\vert^2 - \vert \vec E\vert^2$, and $\vec E\cdot \vec B$
(contained in the $\lambda_j$, Eq.~(\ref{eq:ev})), and the
energy density  $(\vert \vec B\vert^2 + \vert \vec E\vert^2)/8\pi$ and Poynting vector $(c/4\pi)\vec E\times \vec B$ components (contained in the normalization factors as exemplified in
Appendix A, Eq.~(\ref{eq:ed})).

Once an exact or approximate expression for $\Psi_D$ is given, a property of high interest is the electron spin $(\hbar/2)\Sigma^{\mu\nu}$. Its evolution, within Dirac formalism, is given by
\begin{equation}
\langle \Psi_D\vert \frac{i}{2}[\Sigma^\mu_\nu, \hat H_D]\vert\Psi_D\rangle = \mu_B  \langle \Psi_D\vert \gamma^0(\Sigma_{\eta\nu} F^{\mu\eta}- \Sigma^{\eta\mu} F_{\nu\eta})\vert\Psi_D\rangle,
\end{equation}
with $\hat H_D$ the Dirac Hamiltonian.
The contribution of the magnetic field to the precession of spin is encoded in the terms with nonzero value of the indices $\mu$,$\nu$ and $\eta$;
BMT expression  \cite{BMT} for the contribution of the electric field depends on the electron velocity and, as a consequence, requires making a semiclassical approximation involving the terms with one   of those indices equal to zero. Notice, nevertheless, that the general structure of BMT equation is directly provided by Dirac formalism.

\section{Exact equations for the $\mathcal{S}^{(j)}$ functions}
Using Eq.~(\ref{eq:psiP}) Pauli equation reads
$$\sum_{j=1}^4 e^{-i\mathcal{S}^{(j)}/\hbar}\Big[\pi_\mu^{(j)}\pi^{(j)\mu} - m^ 2 c^ 2 +i\hbar (\partial^\mu \pi_\mu^{(j)}) + 2i\hbar\pi_\mu^{(j)}\partial^\mu -\hbar^ 2\partial^ \mu\partial_\mu    \Big] d_j\psi_j$$
\begin{equation}
 =\sum_{j=1}^4  e^{-i\mathcal{S}^{(j)}/\hbar}\frac{q\hbar}{2c}\Sigma^ {\mu\nu}F_{\mu\nu}d_j\psi_j =\sum_{j=1}^4  e^{-i\mathcal{S}^{(j)}/\hbar}m\lambda_j\mu_Bd_j\psi_j.
\label{eq:KGP}
\end{equation}
For given values of the $d_j$ functions, this is a  second order differential equation for the $\mathcal{S}^{(j)}$ functions.
For $\lambda = 0$ in a region of space-time $\mathfrak{R}$ of measure different from zero, the Pauli equation is equivalent to a set of four uncoupled equations
for each bispinor $\psi_j$ on $\mathfrak{R}$.

That is not the case for $\lambda_j\ne 0$.  The Dirac-Pauli equation in the basis $\{\psi_j\}$--composed of bispinors with the structure given by Eq.~(\ref{eq:strcture}) -- can  be written as
$$
\sum_{j=1}^4 e^{-i\mathcal{S}^{(j)}/\hbar}\times\Big[(\pi_\mu^{(j)}\pi^{(j)\mu} - m^ 2 c^ 2 - m\mu_B\lambda_j)d_j
+2i\hbar\pi_\mu^{(j)}\mathcal{P}^{(j)\mu}-\hbar^2\mathcal{Q}^{(j)}+ i\hbar d_j \partial^\mu \pi_\mu^{(j)}$$
\begin{equation}+ 2i\hbar (\partial^\mu e^{-i(S^{(i_j)} -S^{(j)})/\hbar})(\Xi_j^\dagger\partial_\mu d_{i_j}\psi_{i_j})\Big] \psi_j = 0  \label{eq:exact_r}\end{equation}
with
\begin{eqnarray}
\pi_\mu^{(j)} &=& \partial_\mu \mathcal{S}^{(j)} -\frac{q}{c}A_\mu \\
\mathcal{P}_\mu^{(j)} &=& \Xi_j^\dagger\partial_\mu d_j\psi_j
+e^{-i(S^{(i_j)} -S^{(j)})/\hbar}\Xi_j^\dagger\partial_\mu  d_{i_j}\psi_{i_j}\\
\mathcal{Q}^{(j)} &=&  \Xi_j^\dagger\partial^\mu\partial_\mu d_j \psi_j\nonumber+ e^{-i(S^{(i_j)} -S^{(j)})/\hbar}\Xi_j^\dagger\partial^\mu\partial_\mu d_{ij} \psi_{i_j}.
\label{eq:definitions}\end{eqnarray}
 Notice that this equation was obtained by exploiting the algebraic completeness relation given in Eq.~(\ref{eq:proj2}) and the structure of the $\psi_j$ bispinors, Eq.~(\ref{eq:spinors}).
Equation~(\ref{eq:exact_r}) is valid at space-time regions where $S^{(i_j)}$  and $S^{(j)}$ are analytic functions.
Since $\{\psi_j\}$ is formed by linear independent bispinors, this equation gives rise to four coupled scalar equations
\begin{equation}
(\pi_\mu^{(j)}\pi^{(j)\mu} - m^ 2 c^ 2 - m\mu_B\lambda_j)d_j
+2i \hbar\pi_\mu^{(j)} \mathcal{P}^{(j)\mu}
-\hbar^2 \mathcal{Q}^{(j)}
+ i\hbar d_j\partial^\mu \pi_\mu^{(j)} + 2i\hbar (\partial^\mu e^{-i(S^{(i_j)} -S^{(j)})/\hbar})(\Xi_j^\dagger\partial_\mu d_{i_j}\psi_{i_j}) = 0 .\label{eq:Sj1}
\end{equation}
These  are second order differential equations for the $\mathcal{S}^{(j)}$ and $d_j$ functions.
If $d_j\ne 0$, these equations can also be written as
$$
[(\pi_\mu^{(j)} + i\hbar \mathcal{P}_\mu^{(j)}/d_j) (\pi^{(j)\mu} + i\hbar \mathcal{P}^{(j)\mu}/d_j)- m^ 2 c^ 2 - m\mu_B\lambda_j]d_j
-\hbar^2\Big(\mathcal{Q}^{(j)} - \frac{\mathcal{P}_\mu^{(j)} \mathcal{P}^{(j)\mu}}{d_j}\Big)$$
\begin{equation}
+ i\hbar d_j \partial^\mu \pi_\mu^{(j)} + 2i\hbar (\partial^\mu e^{-i(S^{(i_j)} -S^{(j)})/\hbar})(\Xi_j^\dagger\partial_\mu d_{i_j}\psi_{i_j}) = 0 .\label{eq:Sj2}
\end{equation}
Equations~(\ref{eq:Sj1}-\ref{eq:Sj2}) show that, in general, the evolution of the electron can be described by equations
that couple only by pairs $\{j,i_{j}\}$ the eigen bispinors $\{ \psi_j\}$. Physical insight to this 
coupling is provided by the $\mathcal{P}_\mu^{(j)}$ vector. Through its first term, this vector incorporates the expected effects  on $S^{(j)}$ 
of a bispinor $\psi_{j}$ that adiabatically follows the EM field;  through its second term, it incorporates
the effects that arise from the breaking of such adiabatic evolution  yielding an inevitable projection to the pair bispinor $\psi_{i_{j}}$. The terms that define the scalar factor $\mathcal{Q}^{(j)}$ exhibit an analogous behavior.

Notice that the exact Eq.~(\ref{eq:Sj2}) can be related to those of a particle with 
an internal degree of freedom with options  $\{j,i_{j}\}$ immersed in an effective
electromagnetic field $A_\mu^{eff}$ with a contribution of the $\mathcal{P}_\mu^{(j)}$
vector, and with an effective mass which depends on the scalars $\mathcal{Q}^{(j)}$
and $\mathcal{P}_\mu^{(j)}\mathcal{P}^{\mu(j)}$. This similarity is exploited in the method described in Section V.B.

\section{Semiclassical approximations} 

The semiclassical approximations for spinless particles are based in series expansions on $\hbar$ taking this fundamental constant to parametrize the quantum character of each term.
We are interested in making an appropriate description of the influence of the electron spin on the dynamical evolution of its position in physical situations where the effects of Heisenberg
uncertainty relation $\Delta x \Delta p \ge \hbar/2$ can be neglected, that is, when it makes sense to apply the concept of trajectory.
The fact that the  spin of the electron is also proportional to $\hbar$ causes the standard procedure of taking the limit $\hbar\rightarrow 0$ to eliminate  the spin effects on the resulting evolution of the trajectories. Notice, however, that there are qualitative criteria for identifying  terms that have their roots on the quantum dynamics of a point particle.

Newton classical equations restrict the relationship between the  momentum and position of the particle. The least action principle, imposes that variations of the momenta like $\partial^\mu \pi_\mu^{(j)}$
should be null along the classical trajectories. Thus, the term proportional to $\partial^\mu \pi_\mu^{(j)}$ in Eq.~(\ref{eq:Sj1}) should be negligible when semiclassical trajectories are used to describe the position dynamics of an electron.

It is also possible to identify in Eq.~(\ref{eq:Sj1}) the term that arises directly from the interaction of the
electron intrinsic magnetic moment with the EM field, which is the term proportional to $\lambda_j$.  As mentioned above, Rubinow and Keller \cite{RBK} neglect this term irrespective of the strength of the EM field $F_{\mu\nu}$; by doing so, any spin effect on the  trajectory of electrons is suppressed.

The remaining terms in Eq.~(\ref{eq:Sj1}) couple the translational factor $\pi_\mu$ to the variations of the normalized $\psi_j$ and $\psi_{i_j}$ bispinors. For a given structure of the EM
field, $F_{\mu\nu}$, the modules of these terms do not increase as the strength of the field increases. As a consequence, their relevance on Pauli equations is directly related to the spacetime variations
of the electric and magnetic fields.
Notice,  that these terms also include effects from {\it quantum superpositions} of bispinors $\psi_j$ and $\psi_{i_j}$ with eingenvalues $\lambda_j$ and $\lambda_{i_j}$. A semiclassical
description cannot fully incorporate their consequences since a given electron trajectory can only have unique values of the translational factor $\pi_\mu$ that depends on the eigenvalue index, $i.$ $e.$, either $j$ or $i_j$.

\subsection{A semiclassical description of an electron in a high intensity electromagnetic field}
If the electron is immersed in an EM field so intense that the term $\mu_B\lambda_j $
dominates the dynamics over the other terms proportional to $\hbar$ and $\hbar^2$ in Eq.~(\ref{eq:Sj1}),
for each of the Pauli bispinors $\psi_j $,
\begin{equation}
\pi_{\mu}^{(j)}\pi^{(j)\mu} - m^ 2 c^ 2 -m\mu_B\lambda_j \sim 0.\label{eq:OPaprox}
\end{equation}

For real vector potentials $ A_\mu$, which is a requirement to fulfill the continuity equation, we can express
$$\pi_\mu^{(j)} =  \eta_\mu^{(j)} + i \zeta_\mu^{(j)},$$
\begin{equation}
\eta_\mu^{(j)} =  \partial_\mu\mathtt{S}^{(j)} - \frac{q}{c} A_\mu, \quad\quad
\zeta_\mu^{(j)} =  \partial_\mu \mathtt{s}^{(j)}.
\end{equation}
Writing $\mu_B\lambda_j$   as
\begin{equation}
\mu_B\lambda_j = \ell_j + i \mathit{l}_j ,\\
\end{equation}
with  $\ell_j$, $\mathit{l}_j$ real functions, yields the coupled equations
\begin{eqnarray}
(\eta^{(j)\mu} \eta^{(j)}_\mu - \zeta^{(j)\mu} \zeta^{(j)}_\mu - m^2 c^2) -m\ell_j&=&0,\label{eq:reHJ}\\
2\eta^{(j)\mu} \zeta^{(j)}_\mu -m\mathit{l}_j&= &0\label{eq:imHJ}.
\end{eqnarray}
Note that, within the  approximation Eq.~(\ref{eq:OPaprox}), $\eta^{(j)}_\mu$ and $\zeta^{(j)}_\mu$ do not depend on the normalization of the bispinor $\psi_j$, nor on the spinors $\alpha_j$;
 $\eta^{(j)}_\mu$ and $\zeta^{(j)}_\mu$ do depend on $\lambda_j$.
Equation (\ref{eq:reHJ}) has the structure of a Hamilton-Jacobi  equation
where the term $\zeta^{(j)\mu} \zeta^{(j)}_\mu$ resembles a relativistic  analog of the quantum pressure term that arises in the Bohm approach to Schr\"odinger equation. Notice, however, that $\zeta^{(j)\mu}\zeta^{(j)}_\mu$  has its origin on the electron magnetic dipole moment, and it is additional to other terms that could arise from the kinetic energy terms neglected in the semiclassical approximation of Eq.~(\ref{eq:KGP}).  In turn, Eq.~(\ref{eq:imHJ}) can be interpreted as a constraint on the structure of the term $\zeta^{(j)}_\mu$; it is required to guarantee the approximate fulfillment of the Dirac equation and the feasibility of the continuity equation.

We now explore the consequences of Eqs.~(\ref{eq:reHJ}-\ref{eq:imHJ}) on  the Newton equation for the semiclassical electron trajectories.
A general procedure corresponds to identifying \begin{equation}\eta_0 = \frac{1}{c} \mathcal{H} - \frac{q}{c}A_0,\quad\quad \eta_i = p_i -\frac{q}{c}A_i\end{equation}
with $\mathcal{H}$ the Hamiltonian and $p_i$ the canonical momentum; Hamilton- Jacobi equation implies
\begin{eqnarray}
\mathcal{H} &=& c\sqrt{\vec \eta\cdot\vec\eta + \tilde{m}_j^2 c^2} + qA_0,\\
\tilde{m}_jc &=& \sqrt{\zeta^\mu \zeta_\mu + m^2 c^2 +m\ell_j}.
\end{eqnarray}
The equations of motion make sense in standard terms when the square roots factors involved in them yield just real valued functions.
The constraint Eq.~(\ref{eq:imHJ}) should be guaranteed as well as Hamilton equations,
\begin{equation} \frac{d x_i}{dct}= \frac{\partial{\mathcal{H}}}{\partial c p^i} ; \quad\quad \frac{d p_i}{dct} =-\frac{\partial{\mathcal{H}}}{\partial c x^i}.  \end{equation}
The description in terms of the proper time of the electron $\tau$ is achieved by demanding,
\begin{equation}
 \dot{x}_\mu\dot{x}^\mu =1,\quad\dot{x}_\mu = \frac{d x_\mu}{dc\tau}.\label{eq:proper}
\end{equation}
Applying Hamilton equations directly one obtains
\begin{eqnarray}
\dot x_\mu &=& \frac{\eta_\mu}{\tilde{m}_jc},\label{eq:numu}\\
\frac{d(\tilde{m}_j \dot x_\mu)}{dc\tau} &=& \frac{q}{c^2} F_{\mu\nu}\dot x^\nu +\frac{\partial\tilde{m}_j }{\partial x^\mu}\label{eq:newton}.
\end{eqnarray}
The vector function $\zeta_\mu$ can be expected to be a combination of the relevant dynamical variables with the proper behavior under Lorentz transformations. These variables are the position $x_\mu$,
the velocity $\dot x_\mu$, and the contractions of the EM tensors with them $F_{\mu\nu}x^\nu$, $F_{\mu\nu}\dot x^\nu$.
The variables that depend explicitly on the position $x_\mu$ and $F_{\mu\nu}x^\nu$ break the intrinsic homogeneity of space; while a term proportional to $F^{\mu\nu}\dot x_\nu$
gives a null contribution to Eq.~(\ref{eq:imHJ}) and  is not consistent with Eq.~(\ref{eq:numu}) since it would yield a velocity dependent mass.
Thus  we make the {\it ansatz}  $\zeta_\mu = \dot x_\mu \kappa $; then,  Eqs.~(\ref{eq:reHJ}-\ref{eq:imHJ}) as well as Eqs.~(\ref{eq:proper}-\ref{eq:numu})
are fulfilled for the classical trajectories if $\kappa =m\mathit{l}_j/2\tilde{m}_j c$, and
\begin{eqnarray}
\zeta_\mu\zeta^\mu &=&\frac{ - (m^2c^2 + m\ell_j)  + \sqrt{(m^2c^2 + m\ell_j)^2 + m^2 \mathit{l}_j^2}}{2}\nonumber\\ &=:&(\Delta M_j)^2 c^2 \ge 0 \label{eq:sim}.
\end{eqnarray}
The final expression for the effective mass that is required to determine the equations of motion that incorporate spin effects
in the evolution of the electron trajectory is
\begin{equation}
\tilde{m}_j^2 c^2 = \frac{ m^2c^2 + m\ell_j + \sqrt{(m^2c^2 + m\ell_j)^2 + m^2 \mathit{l}^2}}{2};
\end{equation}
$\tilde{m}_j$ does not  depend explicitly on the velocities $\dot x_\mu$, even though $\zeta_\mu$ does.
The expression of $\tilde{m}_j$ generalizes the proposal worked out by Barut \cite{barut} for the particular case that there exists a reference frame where
the electric field is null, so that $l_j = 0$,  and $\ell_j$ is determined by the magnetic field $\vec B$, $\ell_\pm =\pm\mu_B \vert \vec B\vert$. In this case,
$\tilde{m}_j^2 c^2 =  m^2c^2 + m\ell_j$, and the effective mass may be smaller than the inertial mass $m$; besides $\zeta_\mu =0$ is a solution of both
Eqs.~(\ref{eq:reHJ}-\ref{eq:imHJ}), and the function $S^{j}$ can be approximated by a real function.

In general, the effective equation
\begin{equation}
\eta^{(j)\mu} \eta^{(j)}_\mu  - \tilde m_j^2 c^2 =0,\label{eq:sre}
\end{equation}
can be obtained as a condition for trajectories that yield extreme values of the action functional,
\begin{equation}
\mathtt{S}_{eff}^{(j)} =\int dc\tau[\tilde{m}_j c\sqrt{\dot{x}_\mu\dot{x}^\mu} -\frac{q}{c}\dot x_\mu A^\mu].\label{eq:action1}
\end{equation}
The functional
\begin{equation}
-\int dc\tau \vert\Delta M_j\vert c\sqrt{\dot{x}_\mu\dot{x}^\mu} ,\label{eq:seff1}
\end{equation}
gives rise to the effective equation
$$
\zeta_\mu^{(j)} \zeta^{(j)\mu} - (\Delta M_j)^2c^2 = 0 \label{eq:sim}
$$
for its extreme trajectories, and could be identified as $\mathtt{s}_{eff}^{(j)}$. This functional is, by construction, negative definite so that it
guarantees a proper behavior of the exponent factor $e^{-i\mathcal{S}^{(j)}/\hbar}$ 
in the expression of $\Psi_P$.
Notice, however, that there is not a classical action principle that supports the extreme trajectories of that functional as physical trajectories in the semiclassical regime: the action principle refers just to $\mathtt{S}_{eff}^{(j)}$.
Taking Eq.~(\ref{eq:seff1}) as the semiclassical functional $\mathtt{s}_{eff}^{(j)}$ must be complemented by demanding
its evaluation on its extreme trajectories, that do not coincide with the
semiclassical trajectories of the electron.

An alternative for finding a $\mathtt{s}_{eff}^{(j)}$ would be to solve the equation
\begin{equation}
\partial_\mu \mathtt{s}_{eff}^{(j)}  = \dot{x}_\mu\frac{m\mathit{l}_j}{2\tilde{m}_j c}= (\partial_\mu \mathtt{S}_{eff}^{(j)} - \frac{q}{c}A_\mu) \frac{m\mathit{l}_j}{2\tilde{m}_j c} \label{eq:action2}
\end{equation}
with the adequate boundary conditions.
This can be accomplished noticing that Eq.~(\ref{eq:action2}) gives rise to a wave equation for $\partial_\mu\partial^\mu \mathtt{s}_{eff}^{(j)}$ with a known source. Up to a solution of the homogeneous equation, a closed expression of
$\mathtt{s}_{eff}^{(j)}$ can  be written using Green functions. Equation (\ref{eq:action2}) should then be demanded. Contrary to the option given by Eq.~(\ref{eq:seff1}), we were not able to find by this method a functional such that  $\mathtt{s}_{eff}^{(j)}<0$ in general.

There are several interesting properties of the effective mass $\tilde m$. If there exist a reference frame on which $\vec B =\vec 0$, $\ell_j = 0$, $l_\pm =\pm\mu_B \vert \vec E\vert$, and $\tilde{m}_j^2 c^2 =  \Big(m^2c^2 + \sqrt{m^4 c^4 + m^2l_j^2}\Big)/2$, and the effective mass is always greater than the inertial mass $m$. Besides, a direct calculation shows that
\begin{equation}
\frac{\partial \tilde{m}_j c^2}{\partial x^\mu} =\frac{1}{2\sqrt{(mc^2 + \ell_j)^2 + \mathit{l}_j^2}}\Big[ \tilde{m}_j c^2\frac{\partial \ell_j}{\partial x^\mu}
+\Big( \frac{m\mathit{l}_j}{2\tilde{m}_j}\Big)\frac{\partial \mathit{l}_j}{\partial x^\mu}\Big].
\end{equation}
So that, if $mc^2\gg \vert \ell_j + i\mathit{l}_j\vert$, then
\begin{equation}
\frac{\partial \tilde{m}_j c^2}{\partial x^\mu} \approx\frac{1}{2}\Big[\frac{\partial \ell_j}{\partial x^\mu} + \frac{\mathit{l}_j}{2mc^2} \frac{\partial \mathit{l}_j}{\partial x^\mu}  \Big];\label{eq:pc}
\end{equation}
notice that $2mc^2$ is the lower bound of the energy required to create an electron-positron pair, and that $\mathit{l}_j$ must be zero if the electric field is zero. Eq.~(\ref{eq:pc}) shows that, out of the pair creation regime, the real part of the eigenvalue of $\Sigma_{\mu\nu}
F^{\mu\nu}$ could dominate the spin effects on the  trajectories of electrons in external EM fields.

Rewritten in terms of the laboratory time, the equations of motion take the form
\begin{equation}
\frac{d^2 x^i}{dt^2} = \frac{1}{\tilde{m}_j \gamma}\Big[\frac{q}{c}F^{i\nu}\frac{d x_\nu}{dt} -\frac{q}{c^2}\frac{d x^i}{dt}\vec E\cdot\frac{d\vec x}{dt}
-\frac{1}{\gamma}\Big(\frac{\partial \tilde{m}_j c^2}{\partial x^i} +\frac{\partial\tilde{m}_j}{\partial t}\frac{d x^i}{dt}\Big)\Big],\quad \gamma^{-2} = 1 - \vert d\vec x /dct\vert^2.
\label{eq:bessel}
\end{equation}
 The derivatives of $\tilde{m}_j$ could be written as logarithmic.

\subsection{A self consistent approach towards the semiclassical incorporation of other spin dependent effects.}

The semiclassical equations obtained in the last section do not depend on either the eigenvectors of the operator $\Sigma_{\mu\nu} F^{\mu\nu}$ or on the coefficients $d_j$
that are essential to any proper, even approximate, determination of the state function. These equations, nevertheless, give rise to approximate semiclassical expressions
for the spin dependent amplitude assigned to each $\psi_j$ via the complex function $S^{(j)} =\mathtt{S}^{(j)} + i \mathtt{s}^{(j)}$, Eq.~(\ref{eq:action1}-\ref{eq:seff1}).
As a consequence, the results shown in the last section, require the existence of sets $\{ \psi_j\}$, $\{d_j\}$ and $\{S^{(j)}\}$ for which the contribution
of the discarded terms of Eq.~(\ref{eq:Sj1}) can be neglected in the semiclassical equations for the electron trajectories, Eq.~(\ref{eq:bessel}).
The range of validity of this semiclassical approximation is conditioned to the lack of relevance of the terms not considered in it.

 As already discussed, the term $i\partial^\mu \pi_\mu^{(j)}$  in Eq.~(\ref{eq:KGP}) reflects Heisenberg uncertainty relations. 
It must be negligible whenever the concept of trajectory is applied to describe the electron mechanical evolution.
This term is the only one in Eq.~(\ref{eq:KGP}) that depends on the derivative of the generalized momentum $\pi_\mu$, and, as a consequence,
on the second derivative of $S^{(j)}$, so that it cannot be studied within an effective Hamilton Jacobi formalism.
The real part of $i\partial^\mu \pi_\mu^{(j)}$ involves the imaginary part of $S^{(j)}$, its analog in the non relativistic Schr\"odinger equation
is the standard quantum pressure; the imaginary part of $i\partial^\mu \pi_\mu^{(j)}$ is then related to the continuity equation of the current.

While $i\partial^\mu \pi_\mu^{(j)}$ plays an important role when describing the state function of a Dirac electron, in a trajectory approach it is just required to state the consistency conditions that allow neglecting it.
 For non-relativistic particles and time independent
Hamiltonians it can be shown\cite{landau} that assigning a trajectory to a particle requires that the space dependent de Broglie wavelength $\lambda_{DB} = h/\vert \vec{\eta}\vert$
satisfies $\lambda_{DB} \ll L$, where $L$ corresponds to the natural length that describes the variations of the external field  the particle is immersed in. 
This condition is equivalent to $\vert \vec \eta\vert \gg h/L =\hbar (2\pi/L)=: \hbar\vert \vec k\vert$. A strong analogous condition in the relativistic counterpart
could be stated as $\eta_\mu \gg \hbar k_\mu$ with  $k_0 = 2\pi/cT$ and $T$ the natural time interval related to variations of the field.
The main goal of the current section is to show that, within these conditions, it is possible to estimate the relevance of the terms  in Eq.~(\ref{eq:KGP}) that can be encompassed within an effective Hamilton Jacobi description by incorporating them in semiclassical Newton equations. The proposed method searches for self consistency, and it is described by the following algorithm:
\begin{itemize}
\item [(i)] For a given electromagnetic field the eigenvalues $\lambda_j$ of the operator   $\Sigma_{\mu\nu} F^{\mu\nu}$ are evaluated; this yields both
explicit expressions for the equations of motion, Eq.~(\ref{eq:bessel}), and analytical  approximate functions $\{S^{(j)}\}$,  Eqs.(\ref{eq:action1}-\ref{eq:seff1}) \cite{note}. 
\item[(ii)] The equations of motion are solved, in an analytical way -- whenever it is possible -- or numerically. If $\lambda_j\ne 0$ the relevance of the contribution
of $\mu_B\Sigma_{\mu\nu} F^{\mu\nu}$ as compared to the Lorentz interaction can then be estimated.
\item[(iii)] Taking into account the symmetries of the system under consideration, explicit expressions for the spinors $\alpha_j$ that define the bispinors $\psi_j$ can be proposed.

\item[(iv)]
For $\lambda_j\ne 0$, the factors $\Xi_j^\dagger\partial_\mu\psi_j$ and $\Xi_j^\dagger\partial_\mu\psi_{i_j}$ -- which were originally neglected from Eq.~(\ref{eq:Sj1})-- can now be evaluated.
At this stage, the approximate expressions of $\{S^{(j)}\}$, Eq.(\ref{eq:action1}-\ref{eq:seff1}), can be used to estimate $ \mathcal{P}_\mu^{(j)}$ and  $\mathcal{Q}^{(j)}$.
The contribution of  $ \mathcal{P}_\mu^{(j)}$ and  $\mathcal{Q}^{(j)}$ to the semiclassical equations of motion can be estimated using  the formalism developed in Section IV. According to Eq.~(\ref{eq:Sj2}), if $d_j\ne 0$, it is just necessary to make the replacements,
\begin{eqnarray}
\eta_\mu^{(j)} &\rightarrow&  \partial_\mu\mathtt{S}^{(j)} - \frac{q}{c} A_\mu + \mathfrak{Re}(i\hbar \mathcal{P}_\mu^{(j)}/d_j),  \label{eq:iteration1}\\
\zeta_\mu^{(j)} &\rightarrow&  \partial_\mu \mathtt{s}^{(j)} + \mathfrak{Im}(i\hbar \mathcal{P}_\mu^{(j)}/d_j), \label{eq:iteration2}\\
\ell_j &\rightarrow& \mathfrak{Re}(\mu_B\lambda_j - \hbar^2(\mathcal{Q}^{(j)} - \mathcal{P}^{(j)\mu}  \mathcal{P}_\mu^{(j)} /d_j) ), \label{eq:iteration3}\\
\it{l}_j &\rightarrow& \mathfrak{Im}(\mu_B\lambda_j - \hbar^2   (\mathcal{Q}^{(j)} - \mathcal{P}^{(j)\mu}   \mathcal{P}_\mu^{(j)}/d_j )). \label{eq:iteration4}
\end{eqnarray}
Note that the resulting Newton equations couple the trajectories assigned to $j$ and $i_j$.
At this first iteration  stage we may assume  $d_j$ as constant.
Equation (\ref{eq:iteration1}) shows that the spin-orbit term $i\hbar\pi_\mu\partial^\mu \psi_j$ can give rise to effective electromagnetic fields with an strength determined by $\hbar$ and the scale of variations of the field (encoded in $\psi_j$).
\item[(v)] If the equations of motion obtained from (iv) do not yield significant modifications with respect to those found in (ii), one can consider that the bispinors $\psi_j$  proposed in (iii), and the functions $\{S^{(j)}\}$ used in (iv) describe the evolution of the electron within a trajectory limit. It just remains to find coefficients $d_j$ from boundary conditions and normalization of the Dirac bispinor.
Otherwise, a new approximation to the $\{S^{(j)}\}$ functions are obtained from the substitution of  Eqs.~(\ref{eq:iteration1}-\ref{eq:iteration4}) in Eqs.(\ref{eq:action1}-\ref{eq:seff1}). They can be used to get  new expressions for Eqs.~(\ref{eq:iteration1}-\ref{eq:iteration4}), and the process is repeated. Notice that the set $\psi_j$ can also be modified in the intermediate steps of looking for
self  consistency; then, the process would be restarted from (iii).
\end{itemize}

 We stress that this semiclassical description does not include effects of either $\partial^\mu \pi_\mu^{(j)}$, or  $2i\hbar(\partial^\mu e^{i(S^{(i_j)} -S^{(j)}/\hbar)})(\Xi_j\partial_\mu d_{i_j}\psi_{i_j})$. Both terms have an unavoidable quantum character; the first one as being directly related to the Heisenberg uncertainty relations ; the second one as being a direct consequence of quantum superposition.
After arriving to  a self consistent description of the Hamilton-Jacobi terms to a given accuracy, a second consistency  criteria within the same accuracy  should be applied to support neglecting the remaining terms. In order to estimate the relevance of $\partial^\mu \pi_\mu^{(j)}$,
it is necesssary to compare the typical de Broglie wavelength of the electron to the natural scale of the variations of the electromagnetic field in the physical conditions under consideration. The relevance of the second term can be estimated by their direct evaluation on the semiclassical trajectories resulting from the above algorithm.

\section{An example: an electron in an electromagnetic wave with optical vortices}

\subsection{ Semiclassical high intensity electromagnetic field approximation}

As a particular application of the above formalism, we consider an electron moving in a Bessel beam. These beams are propagation-invariant \cite{durnin},  may exhibit vortices \cite{arlt,gutierrez}, and  reconstruct following the presence of an obstacle \cite{bouchal}. The electric field of such beams is
\begin{equation}\vec{E} =e^{i\Theta}\Big[
\frac{1}{2k_\bot}\Big((\frac{\omega}{c}\mathcal{E}^{TE} + ik_z\mathcal{E}^{TM})J_{m_z-1}(k_\bot\rho)e^{-i\varphi}\vec e_+ +
 (\frac{\omega}{c}\mathcal{E}^{TE} - ik_z\mathcal{E}^{TM})J_{m_z+1}(k_\bot\rho)e^{i\varphi} \vec e_-\Big)  + \mathcal{E}^{TM}J_{m_z}(k_\bot\rho)\vec e_z
 \Big]\end{equation}
with $\Theta=k_z z + m_z\varphi - \omega t $, $\vec e_\pm = \vec e_x \pm i \vec e_y$, $(\rho,\varphi,z)$ the position vector in standard cylindrical coordinates notation, $(\omega,{\mathbf k}_\bot,k_z)$
the 1+3 wavevector,
 $J_{m_z}$ the first kind Bessel function of order $m_z$, and $\mathcal{E}^{TE}$ and $\mathcal{E}^{TM}$
the amplitudes of the transverse electric (TE) and transverse magnetic (TM) modes \cite{comment}.
 The corresponding magnetic field is obtained by the replacements
$\mathcal{E}^{TE}\rightarrow -\mathcal{E}^{TM}$ and $\mathcal{E}^{TM}\rightarrow \mathcal{E}^{TE}$. As a consequence, the superposition of modes $\mathcal{E}^{TE}$ and $\mathcal{E}^{TM}$ with the same order, amplitude and phase leads to an EM wave satisfying the two cross field conditions: $\vec E\cdot \vec B = 0$ and $\Delta_{BE} = 0$.  In this case, the formalism predicts the absence of spin effects on the trajectories. It is interesting to note that, in such a case, analytic solutions to the Dirac equation for Bessel beams have been obtained \cite{birula}; the corresponding Dirac bispinor is built
precisely from an analytic solution of the Klein-Gordon equation of a spinless particle.

For either pure TE modes or pure TM modes $\vec E\cdot \vec B =0$, though  $\Delta_{BE} \ne 0$; the explicit expression of the eigenvalues $\lambda$ for TE modes is determined by
\begin{eqnarray}
\lambda_{TE}^2 &=& \Delta_{BE}^2\\ &=& \vert \mathcal{E}^{TE}\vert^2\Big[- \frac{1}{4}(J_{m_z-1}+J_{m_z+1})^2\cos^2 \Theta_{TE}- \frac{1}{4}(J_{m_z-1}-J_{m_z+1})^2\sin^2 \Theta_{TE}
+ J_m^2\cos^2\Theta_{TE} \Big].\nonumber
\end{eqnarray}
The expression for TM modes is obtained by replacing $\vert \mathcal{E}^{TE}\vert^2\rightarrow -\vert \mathcal{E}^{TM}\vert^2$ and $\Theta_{TE} \rightarrow \Theta_{TM}$ .
In Fig.~1,  the behavior of $\Delta^2_{BE}$ is illustrated for TE Bessel beams. In order to acquire some intuition on the expectations of spin effects on the electron trajectories, the derivatives of $\Delta^2_{BE}$ with respect to $k_\bot\rho$ and the phase $\Theta$ are also illustrated.

Two constants of motion can be identified as direct generalizations of those resulting  when no spin effects are taken into account \cite{hacyan}, with the replacement of the electron mass $m$ by its effective value $\tilde{m}_j$:
\begin{eqnarray}
\mathcal{L}&=&\tilde{m}_j\gamma\Big(\rho^2 \frac{d\varphi}{dct} - \frac{m_zc}{\omega}\Big) + \frac{q}{k_\bot^2}\Big(\frac{k_z}{\omega}\mathcal{E}^{TM}\partial_\varphi J_{m_z}e^{i\Theta_{TM}}-\mathcal{E}^{TE}\rho\partial_\rho J_{m_z}e^{i\Theta_{TE}}\Big), \nonumber\\
\mathcal{P}&=&\tilde{m}_j\gamma\Big(\frac{dz}{dct} - \frac{ck_z}{\omega}\Big) -i\frac{q}{\omega}\mathcal{E}^{TE}J_{m_z}e^{i\Theta_{TE}}.
\end{eqnarray}

\begin{figure}
\begin{tabular}{@{}c @{}c @{}c @{}c}
\includegraphics[scale=.22]{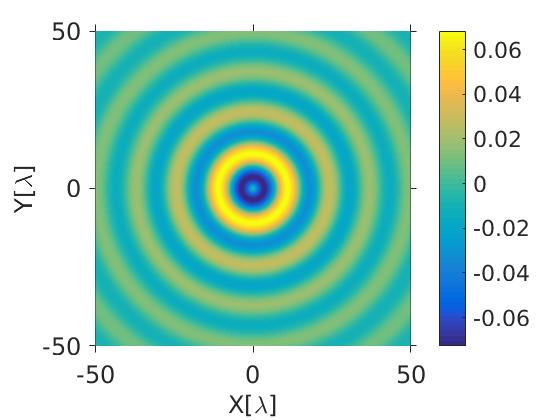}&
\includegraphics[scale=0.22]{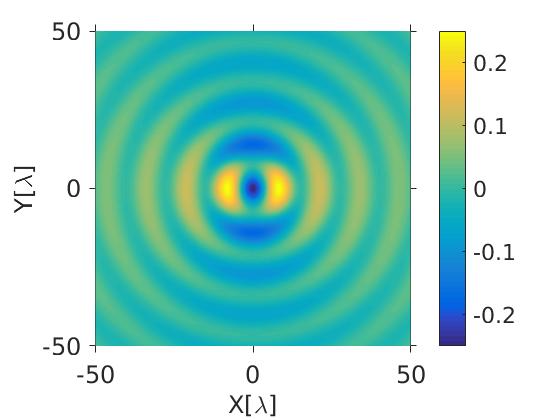}&
\includegraphics[scale=0.22,trim = 0mm 5mm 0mm 0mm]{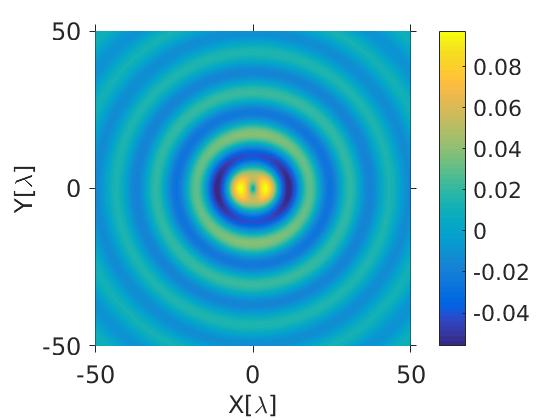}&
\includegraphics[scale=0.22]{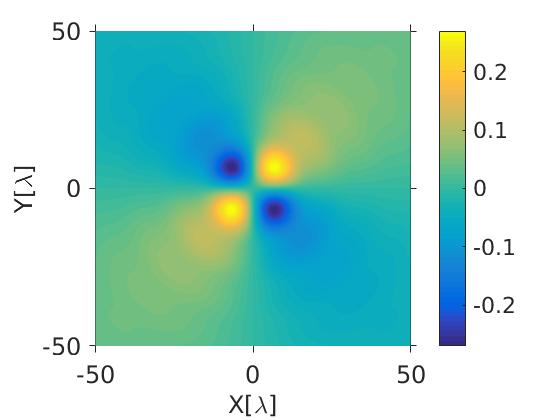}\\
(a) &(b) &(c) & (d)
\end{tabular}
\caption{$\Delta^2_{BE}$ for a TE Bessel mode with $k_\bot = 0.04\times(\omega/c)$ and with (a) $m_z= 0$ and (b) $m_z =1$; (c) radial derivative $\partial_{k_\bot\rho}\Delta_{EB}^2$ and (d) phase derivative $\partial_\Theta\Delta_{EB}^2$ for the $m_z=1$. Results are shown at the plane $k_zz-\omega t =0$ and they are given in units of the squared amplitude $\vert \mathcal{E}^{TE}\vert^2$.\label{fig:DBE}}
\end{figure}

\begin{figure*}
\begin{tabular}{@{}c @{}c @{}c}
\includegraphics[width=.35\textwidth,trim=0mm 50mm 0mm 50mm,]{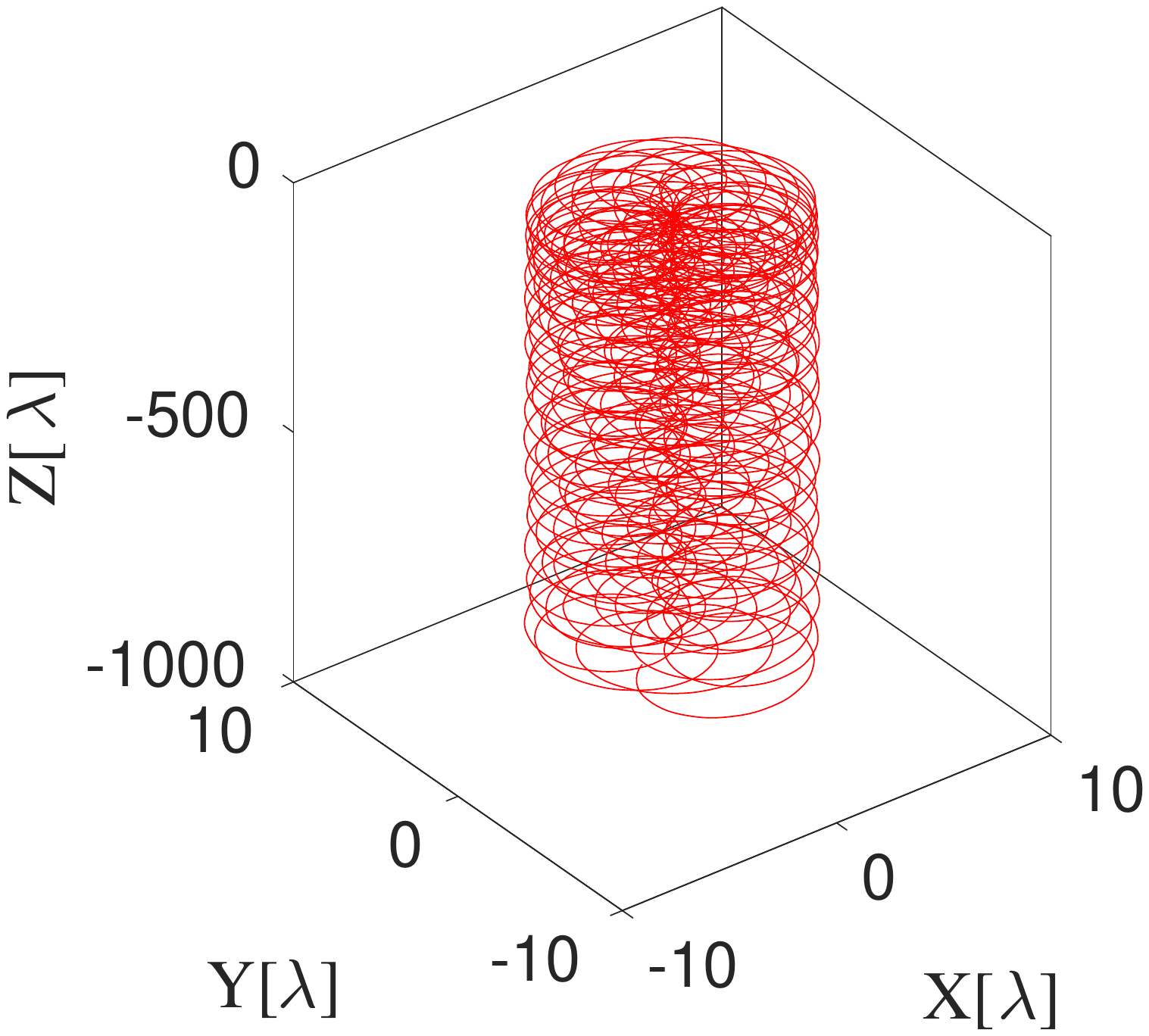} &
\includegraphics[width=0.35\textwidth,trim=0mm 50mm 0mm 150mm,]{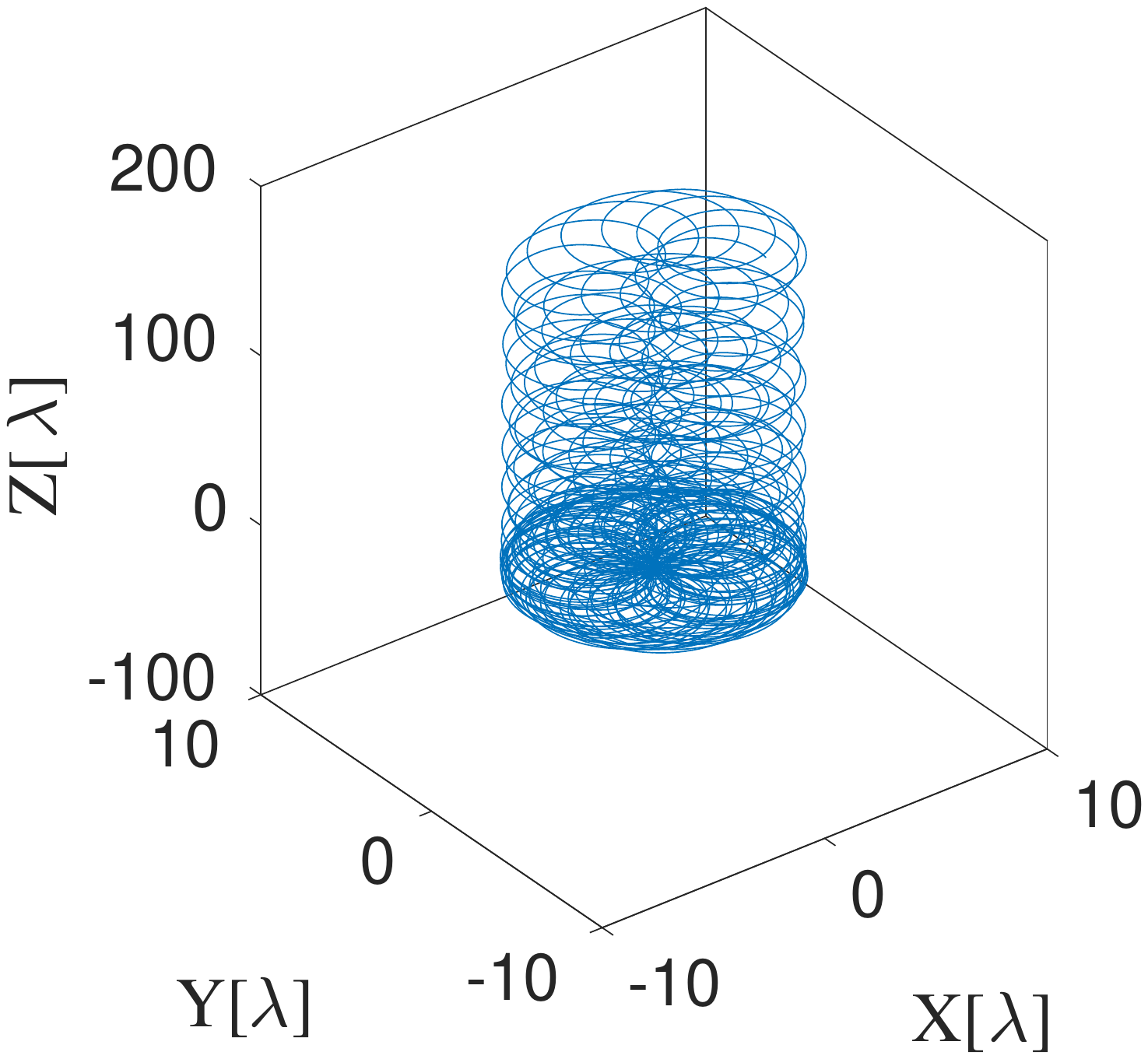}&
\includegraphics[width=0.3\textwidth,trim=0mm 50mm 0mm 50mm,]{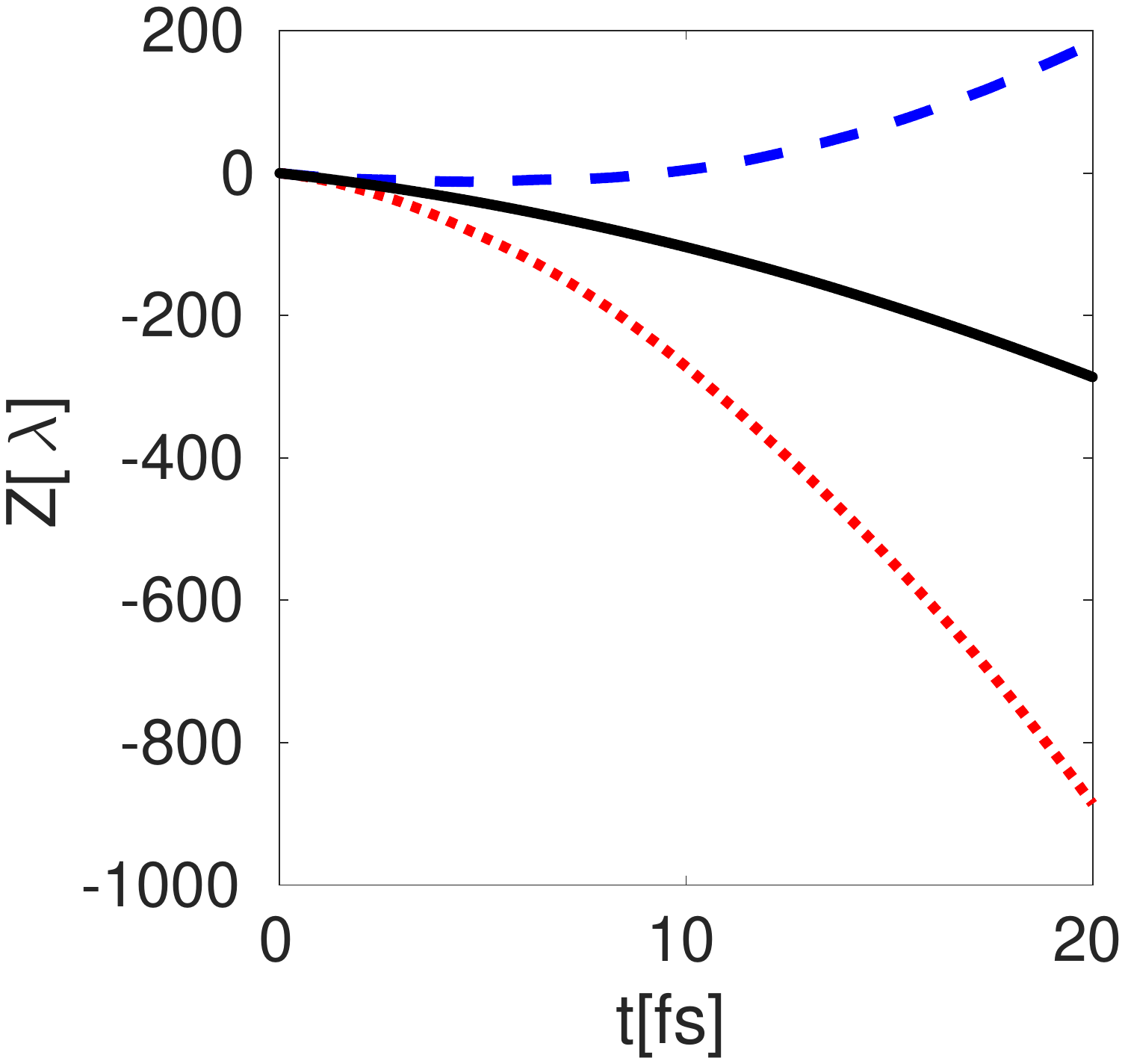}\\
(a) & (b) &  (c)
\end{tabular}
\caption{ Trajectories of an electron in a  $m_z=1$ TE Bessel beam of wavelength 
$\lambda_{beam} =0.1$nm, $k_\bot = 0.04\omega/c$, and $\mathcal{E}^{TE} = 0.005 m c\omega/q$: (a) $\ell = -\mu_B\sqrt{\vert \Delta^2_{BE}\vert}$; (b) $\ell = \mu_B\sqrt{\vert \Delta^2_{BE}\vert}$. (c) $z(t)$ for $\ell = -\mu_B\sqrt{\vert \Delta^2_{BE}\vert}$ (dot line),
$\ell = \mu_B\sqrt{\vert \Delta^2_{BE}\vert}$ (dashed line), and neglecting spin effects (continuous line). The initial conditions are $\rho_0 = 0.05\lambda$, $z_0 =0$, $d\rho_0/dct= 0$, $d\varphi_0/dct =-0.01$, $d z_0/dct = 3\times 10^{-5}$.   }\label{fig:traj2}
\end{figure*}

The effect of spin interactions on an electron trajectory  may be difficult to observe when masked out by the Lorentz force. Even in this scenario, the trajectories of electrons near optical vortices have been proposed as an alternative to helical undulators \cite{birula,hacyan}. In order to enhance spin effects, one can choose
the  length parameters $\delta L$  involved in sensitive variations of  $\Delta_{EB}$ such that $\vert \mu_B\Delta_{EB}/\delta L\vert \sim \vert q F_{\mu\nu}\dot x^\mu \vert$.
For Bessel beams the length parameters are  determined by the inverse of the wave vector components, $k_\bot$ and $k_z$. A direct calculation
shows that a paraxial beam in the  X-ray regime \cite{XrayBessel} is an adequate option. In Fig.~\ref{fig:traj2} we illustrate electron trajectories for X-ray TE Bessel modes ($m_z=1$ and  $\lambda_{beam} = 0.1$nm). The amplitude of the Bessel mode was chosen so as to lead to transverse trapping of the electron; this amplitude is usually expressed in terms of the parameter $e_0 = qE_0/mc$ which is a measure of the coupling of the EM field as compared with the inertia of the charged particle, it has units of frequency. In the simulation we considered $e_0/\omega = 0.005$.   As it usually happens with ultra high intensity laser fields \cite{piazza}, the electron immersed in them rapidly acquires relativistic velocities giving rise -- in the example under consideration-- to de Broglie wavelengths
satisfying  $\lambda_{dB}/2\pi \sim 10^{-12}m \ll \lambda_{beam}$. The latter relation supports the usage of trajectories to describe the electron translational dynamics.
 In this example, conditions are optimized to show that the spin effect can be so large that, depending on the sign of the eigenvalue $\ell_\pm =\pm\mu_B\sqrt{\vert \Delta^2_{BE}\vert}$, there is a directly observable difference on the trajectories: the electron  moves either towards positive or negative values of $z$.

\subsection{Structure of Dirac-Pauli function for TE Bessel modes.}
As described in section V.B, the validity of the semiclassical approximation applied in last subsection, is subject to the effect of the terms discarded from the exact Dirac-Pauli equation Eq.~(\ref{eq:Sj2}). It must be shown that there is an approximate wave function where these terms do not alter, significantly, the semiclassical trajectories for a given $\lambda_j$.
In the Appendices A and B specific basis sets $\{\psi_j\}$ are given and their main properties are described. Appendix B treats the case where the electron evolves in spacetime regions that include surfaces at which $\lambda=0$;
a condition characteristic of the TE modes, as illustrated in Fig.~\ref{fig:DBE}.
In Appendices B and C the main features of the $\{\psi_j\}$ basis --including the representation of the derivatives $\{\partial_\mu \psi_j\}$ and $\{\partial_\mu\partial^\mu \psi_j\}$ on the basis pairs $\psi_j$ and $\psi_{i_j}$ -- are
summarized for TE Bessel modes. In particular, it is shown that
the coupling induced by the Dirac-Pauli equation between the bispinors $\psi_1$ and $\psi_2$ -- which correspond to the eigenvalues
$\pm \sqrt{(\vec B - i\vec E)\cdot(\vec B - i \vec E)}=\pm \lambda_-$ -- is an essential feature of the system. This fact is evident
on the structure of the vectors $\mathcal{P}_\mu$ of Eq.~(\ref{eq:Sj2}),
\begin{eqnarray}
\mathcal{P}_\mu^{(1)} &=& i\hbar d_1\Big[\partial_\mu\Big[\log\frac{a_1\lambda_-d_1}{N_1}\Big] + \frac{G^-_-}{2\lambda_-}\Big[\partial_\mu\Big[\frac{\lambda_- + G^-_z}{G̣_-^-}\Big]\Big] 
+ \frac{d_2}{d_1}\frac{N_1}{N_2}\frac{G^-_-}{2\lambda_-}\Big[\partial_\mu\Big[\frac{\lambda_- - G^-_z}{G̣_-^-}\Big]\Big] e^{-i(\mathcal{S}_2 -
\mathcal{S}_1)\hbar}\Big],\\
\mathcal{P}_\mu^{(2)} &=& i\hbar d_2\Big[\partial_\mu\Big[\log\frac{a_1\lambda_-d_2}{N_2}\Big] + \frac{G^-_-}{2\lambda_-}\Big[\partial_\mu\Big[\frac{\lambda_- - G^-_z}{G̣_-^-}\Big]\Big]
+ \frac{d_1}{d_2}\frac{N_2}{N_1}\frac{G^-_-}{2\lambda_-}\Big[\partial_\mu\Big[\frac{\lambda_- + G^-_z}{G̣_-^-}\Big]\Big] e^{-i(\mathcal{S}_1 -
\mathcal{S}_2)\hbar}\Big].
\end{eqnarray}
In these equations $N_{1,2}(x)$ are normalization factors that guarantee $\psi_j ^\dagger \psi_j =1$, $j=1,2$, and $a_1(x)$ is a complex function that determines the Bloch vector or,
 equivalently, the elementary spinors $\alpha$. Their explicit expressions are given in Appendix C, for Bessel TE modes. Similar expressions are
found for $\psi_3$ and $\psi_4$, which correspond to the eigenvalues
$\pm \sqrt{(\vec B + i\vec E)\cdot(\vec B + i \vec E)}=\pm\lambda_+$.

The vectors $\mathcal{P}^{(j)}_\mu$ give rise to  effective electromagnetic potentials $A^{(j)eff}_\mu$,
 and add  derivative factors to the imaginary part of the action $\mathtt{s}^{(j)}$, Eqs.~(\ref{eq:iteration1}-\ref{eq:iteration2}). The effective potential $A_\mu^{(j)eff}$ yields an effective electromagnetic tensor $F_{\nu\mu}^{(j)eff}=\partial_\nu A^{(j)eff}_\mu -\partial_\mu A^{(j)eff}_\nu$. The effects on the trajectories for each $\lambda_j$ are minimized by demanding $d_1N_2 =d_2N_1$ as can be seen from the resulting equations
\begin{eqnarray}
\mathcal{P}_\mu^{(1)} &=& i\hbar d_1 \Big[\partial_\mu\Big[\log\frac{a_1(\lambda_{-})^2d_1}{N_1G_-^-}\Big] + \Big[e^{-i(\mathcal{S}_1 -
\mathcal{S}_2)\hbar} -1\Big]\frac{G^-_-}{2\lambda_-}\Big[\partial_\mu\Big[\frac{\lambda_- - G^-_z}{G_-^-}\Big]\Big]\Big], \nonumber \\
\mathcal{P}_\mu^{(2)} &=& i\hbar d_2 \Big[\partial_\mu\Big[\log\frac{a_1(\lambda_{-})^2d_2}{N_2G_-^-}\Big] + \Big[e^{-i(\mathcal{S}_2 -
\mathcal{S}_1)\hbar} -1\Big]\frac{G^-_-}{2\lambda_-}\Big[\partial_\mu\Big[\frac{\lambda_- + G^-_z}{G_-^-}\Big]\Big]\Big]. \label{eq:PS} \end{eqnarray}
At the null surfaces, where $\lambda =0$, the second terms in the expression of $\mathcal{P}_\mu^{(j)}$ are singular unless $\mathcal{S}_1 =\mathcal{S}_2$ at those surfaces.

The contribution of the first terms  to  $F^{(j)eff}_{\mu\nu}$ result from taking their real part and it is identically zero
since they are equivalent to a gauge transformation. The imaginary part of these terms
modify the amplitude of the $\psi_j$  contribution to the wave function that would be assigned for the electron:
\begin{eqnarray}
\Psi_D &=& [\gamma^\mu \hat\pi_\mu +mc]\Psi_P\\
\Psi_P &=& N_P[d_1e^{-i\mathcal{S}_1/\hbar}\psi_1 +d_2e^{-i\mathcal{S}_2/\hbar}\psi_2]\nonumber\\
&=& d_1\Big(e^{-i\mathcal{S}_1/\hbar}\psi_1
- \frac{N_2}{N_1}e^{-i\mathcal{S}_2/\hbar}\psi_2),
\end{eqnarray}
\begin{eqnarray}
\mathcal{S}^{(j)} &\sim& \mathtt{S}^{(j)}_{sc} + i \mathtt{s}^{(j)}_{sc},\\
\mathtt{S}^{(j)}_{sc}&=& \int dc\tau\Big[\tilde{m}_j c\sqrt{\dot{x}^\mu\dot{x}_\mu} -\dot x^\mu \Big[\frac{q}{c}A_\mu - \hbar A^{(j)eff}_\mu\Big]\Big]\nonumber,\\
\mathtt{s}^{(j)}_{sc}&=&-\int dc\tau \vert \Delta M_jc\vert\sqrt{\dot x_\mu\dot x^\mu}-\hbar \log\Big[\frac{\vert a_1(\lambda_-)^2\vert}{\vert N_j G_-^-\vert}\Big],\nonumber
\end{eqnarray}
with $N_P$ a normalization constant.
Substitution of the explicit expressions of $\psi_{1,2}$ yields
\begin{eqnarray}
\Psi_P &\sim & N_P\begin{pmatrix} \phi_P \\ \phi_P\end{pmatrix},\\
\phi_P &=& \begin{pmatrix} \mathcal{Y}_-G_-^-/\lambda_-  \\-\mathcal{Y}_+ -\mathcal{Y}_-G_z/\lambda_-\end{pmatrix},\\
\Delta \mathcal{Y}_\pm &=& e^{-i\mathcal{S}^{(1)}_{eff}} \pm e^{-i\mathcal{S}^{(2)}_{eff}},\\
\mathcal{S}^{(j)}_{eff} &=&  \int dc\tau\Big[\tilde{m}_j c\sqrt{\dot{x}^\mu\dot{x}_\mu} -\dot x^\mu \frac{q}{c}A_\mu \Big]\nonumber\\
&-& i \int dc\tau \vert \Delta M_jc\vert\sqrt{\dot x_\mu\dot x^\mu}.
\label{eq:explicit}
\end{eqnarray}
${S}^{(j)}_{eff}$, depends on the index $j=1,2$ through the local eigenvalues $\lambda_1(x) = - \lambda^2(x)$, thus, it is expected that
${S}^{(1)}_{eff}= {S}^{(2)}_{eff}$ at $\lambda_1(x) = \lambda^2(x) = 0$. Notice that for the system under consideration $\Delta M_j$ is zero in regions where
$\lambda_j$ is  real  and it is independent of the sign of $\lambda_j$ when it is purely 
imaginary.

The explicit expression of $\Psi_P$ can be used to estimate the modified semiclassical trajectories; this calculation involves solving the coupled equations for $j$ and $i_j$ electrons.
An interpretation of this procedure is given as follows. The wave function given by Eq.~(\ref{eq:explicit}) describes an ensemble of electrons
with either $\Sigma_{\mu\nu}F^{\mu\nu} = \pm \mu_B\lambda_-$, that is, with an internal state $\hat{\mathcal{O}}_D\psi_1$ or $\hat{\mathcal{O}}_D\psi_2$.
An approximate description of the electron dynamics in terms of semiclassical trajectories for an electron in a $j$-state
is given by the Newton equations of a particle with effective mass $\tilde {m}_j$. This approximate description will not
be able to describe neither the effects of the quantum Heisenberg uncertainty relations nor the effects of interference terms
arising from the quantum superposition of $\psi_{1,2}$. However, in extreme conditions --as those illustrated in Figure 2 for a TE Bessel mode--
there will be observable consequences on the electron transport when it is immersed in  that external EM field.
In order to give a  numerical support to the latter assessment, we have performed a simulation to estimate the effects of the second term in $\mathcal{P}^{(j)}_\mu$,
Eq.(\ref{eq:PS}) on
the semiclassical trajectories. This simulation considered different initial positions of the electron and evaluated $\mathcal{S}^{(j)}_{eff}$ along them.
The  high localization of the effective fields $F_{\mu\nu}^{(j)eff}$, illustrated in Appendix C, made necessary the incorporation of its
regularization at the boundary of the null surfaces. We performed it via the replacement of the divergent factor $(\lambda_-/\vert E^{TE}\vert)^{-3}$
by  $[\tanh(\vert \lambda_-/\vert E^{TE}\vert)^4] (\lambda_-/\vert E^{TE}\vert)^{-3}$. This condition is consistent with the expected identity $\mathcal{S}^{(1)}_{eff}
=\mathcal{S}^{(2)}_{eff}$ at those surfaces.
The numerical simulations gave negligible corrections to the semiclassical trajectories  reported in last section.

\section{Discussion.}
In this article we have identified  basis sets of Dirac bispinors that seem to be natural for the description of electrons in the presence of
an arbitrary electromagnetic field. These basis are built from eigenvectors of the non Hermitean operator $\Sigma_{\mu\nu}F^{\mu\nu}$, so that
the bispinors follow adiabatically the EM field. The normalization and structure of the bispinors incorporate
properties of the EM field like its relativistic invariants, its  energy density and the Poynting vector; in fact the explicit expression of the Dirac bispinors depends directly on Maxwell equations.
Using these basis sets, the Dirac-Pauli equation is decomposed into two equations that couple just two of the four elements of the basis set.

The above formalism has been used to study a semiclassical approach to Dirac equation. This analysis is not devoted to get exact solutions, but
deals with the several conditions required to make an approximate description of the dynamics of a Dirac electron in terms of trajectories.
To a first approximation, the position of a high-energy electron in an arbitrary EM field is expected to follow semiclassical worldlines.
The standard approach assumes that those trajectories can be described  just by considering the Lorentz force. We have presented a formalism that
incorporates the effect of the spin on the trajectories and singled out scenarios where these effects can be neglected.

The Dirac-Pauli equation has a structure that allows the identification of different terms necessary to built the quantum dynamics of a particle with internal degrees of freedom.
The terms with Hamilton-Jacobi reminiscences, link directly to semiclassical treatments; the spin factor $\mu_B\Sigma_{\mu\nu}F^{\mu\nu}$, that scales
with the relativistic invariants of the EM field, allows the generalization of the concept of spin projection along the EM field; the spin-orbit
factor $i\hbar\pi_\mu\partial^\mu\Psi_P$, with no direct dependence on the intensity of the magnetic and electric fields, makes evident the relevance of the EM  space-time variations
 on the electron dynamics; the $i\hbar \partial^\mu\pi_\mu$ term is a direct consequence of Heisenberg uncertainty relation; and finally
the $\hbar^2\partial_\mu\partial^\mu\Psi_P$ factor, which is also not proportional to the field strength, modifies directly the effective mass of the electron.

We have shown that for intense fields the spin effects on the electron trajectory may dominate over other terms (dependent on powers of $\hbar$ but independent of the amplitude of the EM field)
and can not always be discarded in the $\hbar \rightarrow 0$ limit  taken in conventional methods. However, high intensity fields also give rise to a strong Lorentz force; thus, in order to observe directly spin effects moderate space-time variations of the EM fields must be allowed. Notice that, too strong space-time
variations could make necessary a QED treatment to describe pair creation effects.

We test the validity of the presented approximate semiclassical formalism through a self-consistent method.
 Under the semiclassical approximation the trajectories are seen to satisfy a constrained Hamilton-Jacobi equation: the electron spin affects its trajectories through an effective mass, and the constraint deters the particle from reaching classically forbidden regions. A behavior reminiscent of the quantum pressure term in the Bohmian interpretation of quantum mechanics.
The approximate Dirac wave function is obtained from the approximate self-consistent Pauli wavefunction.

Being derived from a first-principles approach, the adopted scheme elucidates the role of the cross field conditions on the spin effects in the quantum dynamics of a Dirac electron. It allows us to depart from ordinary configurations, where the effect of the spin is veiled by the Lorentz force, and present an explicit configuration where the spin will have a measurable effect on the semiclassical trajectories, {\textit{i. e.}}, a Stern-Gerlach scheme for electrons. 

\section*{Appendix A}
In this Appendix, explicit expressions are given for sets of bispinors $\{\psi_j\}$ that, besides satisfying the eigenvalue Eq.~(\ref{eq:psiq}),
\begin{equation}
\Sigma^ {\mu\nu}F_{\mu\nu}\psi_j = \lambda_j(x)\psi_j.
\end{equation}
form a basis.
Let us consider the following two spinors,
\begin{equation}
\alpha = \begin{pmatrix}e^{-i\phi_B/2}\cos\theta_B/2\\e^{i\phi_B/2}\sin\theta_B/2\end{pmatrix}, \quad \beta =\begin{pmatrix}-e^{-i\phi_B/2}\sin\theta_B/2\\e^{i\phi_B/2}\cos\theta_B/2\end{pmatrix}.\label{eq:Bloch}
\end{equation}
They are orthogonal to each other
\begin{equation}
\alpha^\dagger \beta =0,
\end{equation}
and yield the following vectors in Bloch sphere
\begin{eqnarray}
\alpha^\dagger\vec\sigma\alpha &=&(\sin \theta_B\cos\phi_B,\sin\theta_B\sin\phi_B,\cos\theta_B)=:\vec n, \label{eq:n}\\
\beta^\dagger\vec\sigma\beta &=&-\vec n ,\label{eq:mn}\\
\alpha^\dagger\vec\sigma\beta &=& (\cos\theta_B\cos\phi_B + i \sin\phi_B,-\cos\theta_B\sin\phi_B - i\cos\phi_B,-\sin\theta_B) =:\vec g. \label{eq:g}
\end{eqnarray}
Notice that $(\theta_B,\phi_B)$ may depend on the spacetime coordinates $x_\mu$.

In the particular case $\lambda =0$, the bispinors
\begin{eqnarray}
\psi^0_1 &=&\frac{1}{N}\begin{pmatrix}\vec \sigma\cdot \vec B \alpha\\-i\vec\sigma\cdot \vec E \alpha\end{pmatrix}, \quad
\psi^0_2 =\frac{1}{N}\begin{pmatrix}\vec \sigma\cdot \vec B \beta\\-i\vec\sigma\cdot \vec E \beta\end{pmatrix}, \nonumber\\
\psi^0_3 &=&\frac{1}{N}\begin{pmatrix}-i \vec \sigma\cdot \vec E \alpha\\\vec\sigma\cdot \vec B \alpha\end{pmatrix}, \quad
\psi^0_4 =\frac{1}{N}\begin{pmatrix}-i \vec \sigma\cdot \vec E \beta\\\vec\sigma\cdot \vec B \beta\end{pmatrix}, \\
\end{eqnarray}
satisfy
\begin{equation}
\Sigma_{\mu\nu}F^{\mu\nu} \psi^0_j = 0 ,\quad \psi^{0\dagger}_j\psi^0_i = \delta_{ij},
\end{equation}
when
$$ N^2 = \vert \vec B\vert^2 +\vert \vec E\vert^2. $$
The normalization factor is determined by the energy density of the EM field $(1/8\pi)(\vert \vec B\vert^2 +\vert \vec E\vert^2)$.
The set $\{\psi^0_j,j =1,2,3,4\}$ is a basis.

In the case where $\lambda_j \ne 0$, the  bispinors
\begin{eqnarray}
\psi_1 &=&\frac{1}{N_1}\begin{pmatrix} (\lambda_- - \vec\sigma\cdot\vec G^{-})\alpha\\ (\lambda_- - \vec\sigma\cdot\vec G^{-})\alpha\end{pmatrix},\nonumber\\
\psi_2  &=&\frac{1}{N_2}\begin{pmatrix} (\lambda_- +\vec\sigma\cdot\vec G^{-})\beta\\ (\lambda_- + \vec\sigma\cdot\vec G^{-})\beta\end{pmatrix},\nonumber\\
\psi_3 &=&\frac{1}{N_3}\begin{pmatrix} (\lambda_+ - \vec\sigma\cdot\vec G^{+})\alpha\\-(\lambda_+ - \vec\sigma\cdot\vec G^{+})\alpha\end{pmatrix},\nonumber\\
\psi_4 &=&\frac{1}{N_4}\begin{pmatrix} (\lambda_+ + \vec\sigma\cdot\vec G^{+})\beta\\-(\lambda_+ + \vec\sigma\cdot\vec G^{+})\beta\end{pmatrix},\label{eq:basis}
\end{eqnarray}
satisfy the eigenvalue equations
\begin{eqnarray}
\Sigma_{\mu\nu}F^{\mu\nu} \psi_1 &=& \lambda_-\psi_1, \quad \Sigma_{\mu\nu}F^{\mu\nu} \psi_2 = -\lambda_-\psi_2, \nonumber\\
\Sigma_{\mu\nu}F^{\mu\nu} \psi_3 &=& \lambda_+\psi_3, \quad \Sigma_{\mu\nu}F^{\mu\nu} \psi_4 = -\lambda_+\psi_4,
\end{eqnarray}
$\lambda_{\pm 2} = \vec G^\pm\cdot\vec G^\pm$, so that $\lambda_1 = -\lambda_2$ and  $\lambda_3 =- \lambda_4$.
 The normalization factors $N_j$ are chosen to guarantee that $$ \psi_j^\dagger\psi_j =1;$$ they are found to depend directly on the EM  energy density and on the Poynting vector
along the direction of the Bloch vector $\vec n$:
\begin{eqnarray}
N_1^2 &=& 2[\vert\lambda_-\vert ^2 +2\vec n\cdot(\vec B\mathfrak{Re}\lambda_- - \vec E\mathfrak{Im}\lambda_-)+ ( \vert \vec E\vert^2 +\vert \vec B\vert^2 -(\vec E\times \vec B)\cdot \vec n)],\nonumber\\
N_2^2 &=& 2[\vert\lambda_-\vert ^2 +2\vec n\cdot(\vec B\mathfrak{Re}\lambda_- - \vec E\mathfrak{Im}\lambda_-)+ ( \vert \vec E\vert^2 +\vert \vec B\vert^2 +(\vec E\times \vec B)\cdot\vec n)],\nonumber\\
N_3^2 &=& 2[\vert\lambda_+\vert ^2 +2\vec n\cdot (\vec B\mathfrak{Re}\lambda_+ + \vec E\mathfrak{Im}\lambda_+)+ ( \vert \vec E\vert^2 +\vert \vec B\vert^2 +(\vec E\times \vec B)\cdot\vec n)],\nonumber\\
N_4^2 &=& 2[\vert\lambda_+\vert ^2 +2\vec n \cdot (\vec B\mathfrak{Re}\lambda_+ + \vec E\mathfrak{Im}\lambda_+)+ (\vert \vec E\vert^2 +\vert \vec B\vert^2 -(\vec E\times \vec B)\cdot\vec n)].
\label{eq:ed}\end{eqnarray}
The privileged role of the $\vec n$-direction in these equations is a direct result of the selection of the spinors in the defining Eqs.~(\ref{eq:basis}).

Although, these bispinors are not mutually orthogonal,
\begin{eqnarray}
\psi^\dagger_1\psi_2 &=&\frac{2}{N_1N_2}\vec g\cdot\Big[2i(\vec B\mathfrak{Im}\lambda_{-} + \vec E\mathfrak{Re}\lambda_-)+ \vec E\times\vec B\Big],\nonumber\\
\psi^\dagger_3\psi_4 &=&\frac{2}{N_3N_4}\vec g\cdot\Big[2i(\vec B\mathfrak{Im}\lambda_{+} - \vec E\mathfrak{Re}\lambda_+)- \vec E\times\vec B\Big],\nonumber\\
\psi^\dagger_1\psi_3 &=&0,\quad \psi^\dagger_1\psi_4 = 0,\quad
\psi^\dagger_3\psi_2 =0,\quad \psi^\dagger_2\psi_4 =0, \label{eq:northo}
\end{eqnarray}
they are linear independent whenever $\lambda_\pm\ne 0$ and $\lambda_\pm \ne \mp \vec n\cdot \vec G^\pm$. In the latter case, a similar set can be built, $e.$ $g.$, from eigenspinors of  an alternative Bloch vector $\vec n$, to obtain a linear independent set.

 Notice that given a particular $j$, the bispinor that is not orthogonal to $\psi_j$ is that
with index $i_j$ and  $\lambda_{i_j} =-\lambda_j$. For this pair of indices $\{j,i_j\}$ we define
the bispinors, already mentioned in Eq.~(\ref{eq:proj1}),
\begin{equation}
\Xi_j = \frac{\psi_j - (\psi_{i_j}^\dagger\psi_j)\psi_{i_j}}{ 1 -\vert \psi_{i_j}^\dagger\psi_j\vert^2}. \nonumber
\end{equation}
They satisfy the relations Eq.~(\ref{eq:proj2}),
\begin{equation}
\Xi_j^\dagger\psi_i = \delta_{ij},\quad \sum_{i=1}^4 \psi_i\Xi_i^\dagger = \mathbf{1}_{4\times 4} . \nonumber
\end{equation}

Note also that, in general,
\begin{eqnarray}
\psi^\dagger_1\partial_\mu\psi_3 &=&0,\quad \psi^\dagger_1\partial_\mu\psi_4 = 0,
\psi^\dagger_3\partial_\mu\psi_2 =0,\quad \psi^\dagger_2\partial_\mu\psi_4 =0, \nonumber\\
\psi^\dagger_1\partial_\mu\partial^\mu\psi_3 &=&0,\quad \psi^\dagger_1\partial_\mu\partial^\mu\psi_4 = 0,
\psi^\dagger_3\partial_\mu\partial^\mu\psi_2 =0,\quad \psi^\dagger_2\partial_\mu\partial^\mu\psi_4 =0. \label{eq:overlap}
\end{eqnarray}

\section*{Appendix B}
An specific choice of the spinors $\alpha$ and $\beta$ introduced in Appendix A is necessary to define a particular basis set $\{\psi_j\}$.
For a given physical system, this choice should take into account the symmetries and boundary conditions of that system.
 The electron in the vector Bessel beam studied in Section VI, illustrates a very interesting physical situation where
boundary conditions highly restrict the choice of the basis. In this case, the electron is immersed in a field where the
eigenvalues $\lambda^2_j$ are different from zero in open regions of spacetime but become null at  defined surfaces,
lines or isolated points;  generically, we call these regions ``null surfaces''.
The Dirac wavefunction and its derivatives should be continuous at those low dimension regions. The latter condition imposes
the usage of basis sets $\{\psi_j\}$ with the structure  Eq.~(\ref{eq:basis}) for $\lambda^2\ne 0$ over the whole spacetime,
but with adequate behavior on the null surfaces.
The derivatives of $\lambda$ are directly given by
\begin{equation}
\partial_\mu \lambda = \frac{\vec G\cdot \partial_\mu \vec G}{\lambda},
\end{equation}
which, in general exhibits a divergence as $\lambda \rightarrow 0$. It must be guaranteed that this divergence does not affect the
 behavior of the derivatives of the basis bispinors $\partial_\mu \psi_j$, $j=1,...,4$ at the null surfaces. According to Eq.~(\ref{eq:basis}),
the bispinors $\psi_j$  are determined by spinors $\alpha_j$. It is thus necessary to demand  that the coefficients of $\partial_\mu \lambda_j$
that appear in $\partial_\mu \psi_j$ for given $\alpha_j$ add up to zero as $\lambda_j$ approaches the zero value.
A direct calculation shows that this condition is satisfied if the structure of each $\alpha_j$ is
\begin{equation}
\alpha^{(j)} =\begin{pmatrix} a^{(j)}_1\\ a^{(j)}_2\end{pmatrix}, \quad \quad \frac{a^{(j)}_2}{a^{(j)}_1} =-\frac{G^{(j)}_z}{G^{(j)}_-}, \label{eq:a1a2}
\end{equation}
with $\vec G^{(1)} = \vec B - i \vec E = -\vec G^{(2)} =\vec G^-$ and $\vec G^{(3)} = \vec B + i \vec E = -\vec G^{(4)} = \vec G^+$.
In such a case  the spinors that define $\psi_j$  Eq.~(\ref{eq:spinors}) are
\begin{equation}
\phi^{(j)} = \frac{a^{(j)}_1}{N_j} (\lambda_j - \vec\sigma\cdot\vec G^{(j)})\begin{pmatrix} 1\\ \frac{a^{(j)}_2}{a^{(j)}_1}\end{pmatrix} =  \frac{a^{(j)}_1\lambda_j}{N_j}\begin{pmatrix}  1\\ -\frac{\lambda_j + G^{(j)}_z}{G^{(j)}_-}\end{pmatrix}.\label{eq:phizero}
\end{equation}
Notice that for $\lambda_j\rightarrow 0$, the ratio $\lambda_j/N_j\rightarrow \lambda_j/\vert \lambda_j \vert$ yields a factor  $e^{i\varphi_{\lambda_j}}$ which depends on the trajectory used
to take this limit in the complex plane where $\lambda_j$ is contained.

From this expression a direct calculation shows that the derivative of the bispinor $\psi_j$ can be written as
\begin{equation}
\partial_\mu \psi^{(j)}=(\Xi_j^\dagger\partial_\mu\psi_j)\psi_j + (\Xi_{i_j}^\dagger\partial_\mu\psi_j)\psi_{i_j},
\end{equation}
with
\begin{eqnarray}
\Xi_j^\dagger\partial_\mu\psi_j &=& \partial_\mu[\mathrm{Log}(a^{(j)}_1\lambda_j/N_j)] + \frac{G_-^-}{2\lambda_j}\partial_\mu\Big(\frac{\lambda_j + G_z^{(j)}}{G^{(j)}_-}\Big),\nonumber\\
\Xi_{i_j}^\dagger\partial_\mu\psi_j &=& \frac{G_-^-}{2\lambda_j}\frac{N_{i_j}}{N_j}\Big[\partial_\mu \Big(\frac{\lambda_j + G_z^{(j)}}{G^{(j)}_-}\Big)\Big].
\label{eq:Xis}
\end{eqnarray}
Notice that the expression for $\partial_\mu\partial^\mu \Psi_j$ in the basis $\{\psi_j\}$ can be obtained by taking the second derivative of last equation and make the proper substitution of the first derivatives using again Eq.~(\ref{eq:Xis}).

\section*{Appendix C}

For Bessel modes, the vector $\vec G^{\pm} =\vec B\pm i \vec E$ is given by
\begin{eqnarray}
\vec G^{+}_\pm &=& (B_x \pm i B_y) + i  (E_x \pm i E_y)\nonumber\\
               &=&\frac{\omega }{2k_\bot c}\Big[\pm\frac{k_zc}{\omega} -1\Big] e^{\pm i\varphi}
               \cdot\Big[ \vert{\mathcal E}^{TM}\vert\mathcal{J}^\mp_{m_z}(k_\bot\rho,\Theta_{TM})-i\vert{\mathcal E}^{TE}\vert \mathcal{J}^\mp_{m_z}(k_\bot\rho,\Theta_{TE}) \Big],\label{eq:Gppm}\\
\vec G^{-}_\pm &=& (B_x \pm i B_y) - i  (E_x \pm i E_y)\nonumber\\
               &=&\frac{\omega }{2k_\bot c}\Big[\mp\frac{k_zc}{\omega} -1\Big] e^{\pm i\varphi}
              \cdot\Big[ \vert{\mathcal E}^{TM}\vert\mathcal{J}_{m_z}^\mp(k_\bot\rho,\Theta_{TM})+i\vert{\mathcal E}^{TE}\vert \mathcal{J}_{m_z}^\mp(k_\bot\rho,\Theta_{TE}) \Big], \label{eq:Gmpm}\\
\vec G^{+}_z &=& B_z + i E_z\nonumber\\
             &=& \vert{\mathcal E}^{TE}\vert\cos \Theta_{TE} + i \vert{\mathcal E}^{TM}\vert \cos \Theta_{ TM}\label{eq:Gz},
\end{eqnarray}
with
\begin{eqnarray}
\mathcal{E}^{TM} = \vert\mathcal{E}^{TM}\vert e^{i\theta_{ TM}}&,&
\Theta_{TM} = k_z z + m_z\varphi - \omega t +\theta_{ TM},\nonumber\\
\mathcal{E}^{TE} = \vert\mathcal{E}^{TE}\vert e^{i\theta_{TE}}&,&
\Theta_{TE} = k_z z + m_z\varphi - \omega t +\theta_{ TE},\nonumber\\
\mathcal{J}_m^{\mp}(k_\bot\rho,\Theta) = J_{m-1}&(&k_\bot\rho) e^{\mp i \Theta} + J_{m+1}(k_\bot\rho) e^{\pm i \Theta}.\label{eq:calJ}
\end{eqnarray}
The factor   $\pm k_zc/\omega -1$  in Eqs.~(\ref{eq:Gppm}-\ref{eq:Gmpm}) enhances or suppresses the values of the components $G_{\pm}$ depending on the sign of $k_z$. For a beam
with a given angular momentum along the $Z$-axis, this factor determines  the helicity -- and chirality-- of the Bessel beam. Note also that the phase factor
$\pm\varphi$ in the expressions for $G^\pm$ is a manifestation of the intrinsic vector character of the EM field and the resulting spin angular momentum of light.

In order to check the consistency of the numerical results shown in Figure 2 of Section VI, we need to estimate the factors neglected in the semiclassical approximation described in Section IV for an
electron in a TE Bessel mode. Some of them depend on the factors $\Xi_i\partial_\mu \psi_j$ of the exact equation Eq.~(\ref{eq:Sj1}).
 As the  eigenvalues $\lambda^2_j$ are null on some surfaces, we apply directly the basis  $\{\psi_j\}$ described in Appendix B. These bispinors are given by Eq.~(\ref{eq:phizero}), and depend directly on the spinors $\alpha_j$. The condition Eq.~(\ref{eq:a1a2}) together with the normalization $\vert a_1\vert^2 + \vert a_2\vert^2 =1$ imply that the Bloch angles defined in Eq.~(\ref{eq:Bloch})
satisfy
\begin{eqnarray}
\cos \theta_B &=& \frac{(1+k_zc/\omega)G_z^2 - (1-k_zc/\omega)\vert\mathcal{J}^-_m\vert^2/4}{(1+k_zc/\omega)G_z^2 + (1-k_zc/\omega)\vert\mathcal{J}^-_m\vert^2/4}, \label{eq:cosB}\\
e^{i\phi_B} &=& -i\frac{G_z}{\vert G_z\vert}\frac{\vert G^{-}_-\vert}{G^{-}_-},
\end{eqnarray}
for the  $\alpha_j$ spinors, $j=1,2$. The expressions of the Bloch angles corresponding to the $\alpha_j$ spinors, $j=3,4$ are found by making the replacement  $k_z\rightarrow -k_z$ in the last equation.
The normalization factors of bispinors $\psi_j$ can be evaluated directly from their expression Eq.~(\ref{eq:phizero}) or using the Bloch angles in a similar way as Eq.~(\ref{eq:ed}).

\begin{figure}
\begin{tabular}{@{}c @{}c}
\includegraphics[width=.48\textwidth]{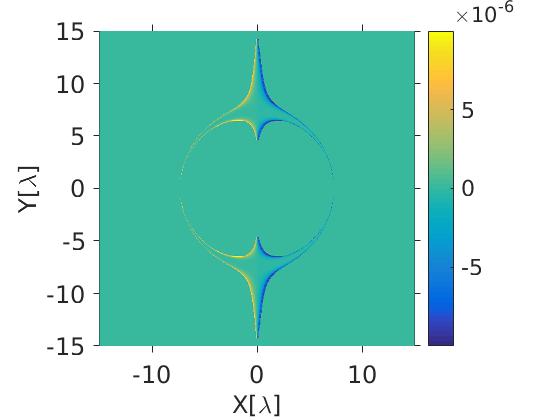} &\includegraphics[width=.48\textwidth]{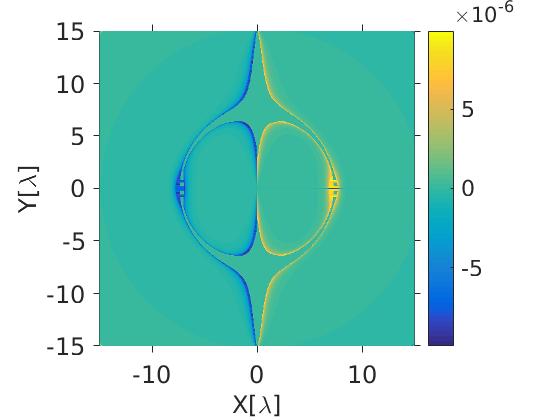}\\
(a) & (b)
\end{tabular}
\caption{(a) Real  and (b) imaginary part of ${\mathcal F}_{z \varphi}^{(j:j)}$ for a TE Bessel beam. Results are shown at the plane $k_zz-\omega t =0$, and the divergence at the null
planes were cut at the value $10^{-5}$ in order to illustrate  the high localization of ${\mathcal F}_{z \varphi}^{(j:j)}$.  }\label{fig:FPT}
\end{figure}

The projections $\Xi_j^\dagger\partial_\mu\psi_{i}$ determine the factor $\mathcal{P}_\mu$, that yield an effective electromagnetic potential via Eq.~(\ref{eq:iteration1}). We need to evaluate the corresponding effective electromagnetic tensor:
\begin{equation}
\frac{q}{c}{\mathcal F}_{\mu\nu}^{(i:j)} =\mathfrak{Re}\Big[i\hbar[ \partial_\nu \Xi_i^\dagger\partial_\mu\psi_{j} - \partial_\mu \Xi_i^\dagger\partial_\nu\psi_{j}]\Big].
\end{equation}
As expected, the amplitude of the effective field does not depend on the amplitude of the Bessel mode. Instead the  fine structure constant $q/\hbar c$ plays a crucial role
in the determination of such an amplitude as shown in last equation.
It is evident that the first term in the expression Eq.(\ref{eq:Xis}), $\partial_\mu[\mathrm{Log}(a^{(j)}_1\lambda_j/N_j)]$ does not contribute to ${\mathcal F}_{\mu\nu}^{(i:j)}$.
After a direct calculation it can be shown that for a TE Bessel mode,
\begin{eqnarray}
\frac{q}{\hbar c}{\mathcal F}_{\mu\nu}^{(j:j)}&=& 2\Big(1- \Big(\frac{k_zc}{\omega}\Big)^2 \Big)\Big[\Big(\partial_\mu \phi_-\Big)\Big(\partial_\nu\Big[\frac{G_z}{\lambda}\Big]\Big)-\Big(\partial_\mu \phi_-\Big)\Big(\partial_\nu\Big[\frac{G_z}{\lambda}\Big]\Big)\Big], \quad e^{i\phi_\pm}= \frac{G^{-}_\pm}{\vert G^{-}_\pm\vert},
\end{eqnarray}
 so that ${\mathcal F}_{\mu\nu}^{(j:j)}$ is real for $\lambda^2<0$, and imaginary for
$\lambda^2>0$.
In Fig.~(\ref{fig:FPT}) we illustrate the behavior of ${\mathcal F}_{z \varphi}^{(j:j)}$.
A general feature is that the derivative of the term $G_z/\lambda$ is singular at $\lambda\rightarrow 0$. It is also very small out of the neighborhood of the null surfaces.

\begin{acknowledgments}
This work was supported by CONACYT LN-280181, M\'exico. R.J. thanks Shahen Hacyan for his interesting comments.
\end{acknowledgments}


\begin{thebibliography} {10}
\bibitem{note}Notice that
 for $\lambda_j =0$, Eq.(\ref{eq:action1}) reduces to the action of a spinless particle-- in accordance to Rubinow and Keller formalism --while the imaginary part $\mathtt{s}^{(j)}$ is null.
\bibitem{comment} Though not written explicitly, the real part of complex expressions give the physical electromagnetic fields.
\end{thebibliography}
\end{document}